\documentclass[useAMS,usenatbib,twocolumn]{mn2e}
\usepackage{amsmath}
\usepackage{amssymb}
\usepackage{graphicx}
\usepackage{color}
\usepackage{amsmath,amsbsy}

\newcommand*\Bell{\ensuremath{\boldsymbol\ell}}
\newcommand{\Ma}{\mathcal{M}}

\title[]{Centroid Velocity Statistics of Molecular Clouds}
\author[Bertram et al.]{Erik~Bertram, Lukas~Konstandin, Rahul~Shetty, Simon~C.~O.~Glover \and Ralf~S.~Klessen\\
Universit\"at Heidelberg, Zentrum f\"ur Astronomie, Institut f\"ur Theoretische  Astrophysik, Albert-Ueberle-Str.~2, 69120 Heidelberg, Germany
}

\begin{document}

\maketitle

\abstract
We compute structure functions and Fourier spectra of 2D centroid velocity (CV) maps in order to study the gas dynamics of typical molecular clouds (MCs) in numerical simulations. We account for a simplified treatment of time-dependent chemistry and the non-isothermal nature of the gas and use a 3D radiative transfer tool to model the CO line emission in a post-processing step. We perform simulations using three different initial mean number densities of $n_{0} = 30, 100$ and $300 \: {\rm cm^{-3}}$ to span a range of typical values for dense gas clouds in the solar neighbourhood. We compute slopes of the centroid velocity increment structure functions (CVISF) and of Fourier spectra for different chemical components: the total density, H$_2$ number density, $^{12}$CO number density as well as the integrated intensity of $^{12}$CO ($J=1 \rightarrow 0$) and $^{13}$CO ($J=1 \rightarrow 0$). We show that optical depth effects can significantly affect the slopes derived for the CVISF, which also leads to different scaling properties for the Fourier spectra. The slopes of CVISF and Fourier spectra for H$_2$ are significantly steeper than those for the different CO tracers, independent of the density and the numerical resolution. This is due to the larger space-filling factor of H$_2$ as it is better able to self-shield in diffuse regions, leading to a larger fractal co-dimension compared to CO.
\endabstract


\section{Introduction}
\label{sec:introduction}

Understanding the dynamics of molecular clouds (MCs) is an important task in astrophysics. It is still a great debate which physical processes regulate star formation. Beyond self-gravity, the radiation field, magnetic fields and a complex thermodynamic and chemical evolution, several studies in the past have shown that turbulent motions play an important role for controlling the star formation process \citep[see, e.g.][]{MacLowAndKlessen2004,ScaloAndElmegreen2004,ElmegreenAndScalo2004,McKeeAndOstriker2007,HennebelleAndFalgarone2012}.

Characterising interstellar turbulence is therefore important for understanding star formation. During the last years large improvements were made in understanding turbulence theory. \citet{Kolmogorov1941} proposed a constant mean energy dissipation rate within an inertial range and assumed that all statistical averaged quantities only depend on this rate, leading to the law of Kolmogorov for incompressible turbulence. This directly leads to the famous 5/3-law of the energy spectrum with a statistically quasi-equilibrium state. Nevertheless, many experimental and theoretical studies in the past decades have shown that the assumption of a constant mean energy dissipation rate is not accurate \citep{BenziEtAl1993,SheAndLeveque1994,Dubrulle1994,Frisch1995} and seems to deviate significantly from measurements for higher orders of the structure function. This discrepancy has been referred to as intermittency corrections to the Kolmogorov theory. A number of phenomenological models have been proposed since then to explain the differences. \citet{SheAndLeveque1994} proposed an intermittent scaling for filamentary structures, in excellent agreement with tunnel-flow experimental data \citep{BenziEtAl1993} for incompressible turbulence. \citet{BoldyrevEtAl2002} expanded the work of \citet{SheAndLeveque1994} for sheet-like shocks or a general fractal dimension of the dissipative structures for compressible turbulence. Later on, other studies developed models for a better understanding of turbulence involving magnetic fields \citep[see, e.g.][]{GoldreichAndSridhar1995,ChoLazarianVishniac2002}.

Studying interstellar turbulence is difficult, since observations are always a complex convolution of the density and the velocity field, affected by many other astrophysical processes. Assessing the influence of projection effects relies on numerical simulations of turbulence that can be compared to observational measurements. In that sense, \citet{LisEtAl1996} have shown that it is possible to use centroid velocity increments (CVI) in order to trace the most intense velocity structures of regions in the interstellar medium (ISM), which do not form stars and to analyse intermittency effects on the basis of the two-point statistics.

Recent studies have revealed in great detail that centroid velocities (CV) and their increments are sensitive to the turbulent driving, density fluctuations and temperatures along the line-of-sight \citep{LazarianAndEsquivel2003,OssenkopfEtAl2006,EsquivelEtAl2007,Hily-BlantEtAl2008,FederrathEtAl2010}. Furthermore, the non-uniform chemical makeup of the gas and optical depth effects might also alter the CVI statistics (see, e.g. \citeauthor{BurkhartEtAl2013b}~2013b and \citealt{BurkhartEtAl2014}). Chemical effects may be important for an accurate determination of the velocity fluctuations, since the relation between H$_2$ as the main constituent of the ISM and CO as a tracer is not linear (see, e.g. \citealt{GloverEtAl2010}; \citeauthor{ShettyEtAl2011a}~2011a). Optical depth effects may as well affect projected structures due to different opacities of the gas tracers (see, e.g. \citeauthor{BurkhartEtAl2013a}~2013a,b, 2014 and \citealt{LazarianEtAl2004}). Nevertheless, the question of how different physical processes affect the overall two-point statistics of the CVI remains elusive although much progress has been made during the last years.

In this paper we analyse high-resolution, three-dimensional chemistry models in numerical simulations of turbulent interstellar gas \citep{GloverEtAl2010} to get a better insight into how the choice of chemical species as a tracer of gas affects the two-point statistics of the CV. Therefore we use time-dependent chemistry for modelling the formation and destruction of molecules in the ISM. We then perform radiative transfer calculations and convert our simulations into synthetic 2D CV maps. We focus in this paper on how chemistry and radiative transfer affect the CVI turbulence statistics and the Fourier spectra in different environments, while spanning a range of possible densities representative of MCs.

In Sec. \ref{sec:methodsandsims} we present the simulations, the methods used to compute the CVI, the Fourier spectra and the radiative transfer post-processing. In Sec. \ref{sec:results} we present the results of our studies with different mean number densities. We discuss our results in Sec. \ref{sec:discussion} and present our conclusions in Sec. \ref{sec:summary}.

\section{Methods and simulations}
\label{sec:methodsandsims}

\subsection{MHD and chemistry simulations}
\label{subsec:sims}

The simulations that we examine in this paper were performed using a modified version of the Z{\sc eus}-M{\sc p} magnetohydrodynamical code \citep{Norman2000,HayesEtAl2006}. We make use of a detailed atomic and molecular cooling function, described in detail in \citet{GloverEtAl2010} and \citet{GloverAndClark2012}, and a simplified treatment of the molecular chemistry of the gas. Our chemical treatment is based on the work of \citet{NelsonAndLanger1999} and \citet{GloverAndMacLow2007}, and allows us to follow the formation and destruction of H$_{2}$ and CO self-consistently within our simulations. Full details of the chemical model can be found in \citet{GloverAndClark2012}.

We consider a periodic volume, with side length 20~pc, filled with uniform density gas. We run simulations with three different initial densities, $n_{0} = 30, 100$ and $300 \: {\rm cm^{-3}}$, where $n_{0}$ is the initial number density of hydrogen nuclei. The initial temperature of the gas is set to 60~K. We initialise the gas with a turbulent velocity field with uniform power between wavenumbers $k = 1$ and $k = 2$ \citep{MacLowKlessenBurkertSmith1998,MacLow1999}. The initial 3D rms velocity is $v_{\rm rms} = 5 \: {\rm km \: s^{-1}}$, and the turbulence is driven so as to maintain $v_{\rm rms}$ at approximately the same value throughout the run. The gas is magnetised, with an initially uniform magnetic field strength $B_{0} = 5.85 \mbox{ }\mu {\rm G}$, initially oriented parallel to the $z$-axis of the simulation. We neglect the effects of self-gravity. We assume that the gas has a uniform solar metallicity and adopt the same ratios of helium, carbon and oxygen to hydrogen as in \citet{GloverEtAl2010} and \citet{GloverAndMacLow2011}. At the start of the simulations, the hydrogen, helium and oxygen are in atomic form, while the carbon is assumed to be in singly ionised form, as C$^{+}$. We also adopt the standard local value for the dust-to-gas ratio of 1:100 \citep{GloverEtAl2010}, and assume that the dust properties do not vary with the gas density. The cosmic ray ionisation rate was set to $\zeta = 10^{-17} \: {\rm s^{-1}}$, but we do not expect our results to be sensitive to this choice \citep{GloverEtAl2010}. For the incident ultraviolet radiation field, we adopt the standard parameterisation of \citet{Draine1978}. More detailed information about the simulations that we examine in this paper can be found in \citet{BertramEtAl2014}. The runs analysed in this paper were all performed with a numerical resolution of $512^3$ zones. We discuss the influence of our limited numerical resolution in more detail in Appendix \ref{app:resolution}.

\subsection{Radiative transfer}
\label{subsec:RADMC}

To model the CO ($J=1 \rightarrow 0$) line in the MCs for both $^{12}$CO and $^{13}$CO we employ the radiative transfer code R{\sc admc}-3{\sc d}\footnote{www.ita.uni-heidelberg.de/$\sim$dullemond/software/radmc-3d/} \citep{Dullemond2012}. We use the Large Velocity Gradient (LVG) approximation \citep{Sobolev1957} to compute the level populations. The LVG implementation in R{\sc admc}-3{\sc d} is described in \citeauthor{ShettyEtAl2011a}~(2011a). For more information about the usage of R{\sc admc}-3{\sc d} on our data we refer the reader to \citet{BertramEtAl2014}.

Our simulations do not explicitly track the abundance of $^{13}$CO, and so we need a procedure to relate the $^{13}$CO number density to that of $^{12}$CO. A common assumption is that the ratio of $^{12}$CO to $^{13}$CO is identical to the elemental abundance ratio of $^{12}$C to $^{13}$C \citep[see, e.g.][]{RomanDuvalEtAl2010}. In the majority of our analysis, we make the same assumption and set the $^{12}$CO to $^{13}$CO ratio to a constant value, $R_{12/13} = 50$.

However, strictly speaking, we expect the $^{12}$CO to $^{13}$CO ratio to faithfully reflect the elemental $^{12}$C to $^{13}$C ratio only in gas where essentially all of the carbon is incorporated into CO. When this is not the case (e.g. at low density and/or low extinction), the effects of chemical fractionation \citep{WatsonEtAl1976} and selective photodissociation of $^{13}$CO \citep[see, e.g.][]{VisserEtAl2009} can significantly alter the $^{12}$CO/$^{13}$CO ratio \citep{RoelligAndOssenkopf2013,SzucsEtAl2014}. To explore the effect that this may have on the statistical properties of the $^{13}$CO line emission, we have produced a set of $^{13}$CO emission maps, where we attempt to account for spatial variations in the $^{12}$CO/$^{13}$CO ratio. To do this, we made use of the recent numerical results of \citet{SzucsEtAl2014}. They find that in turbulent molecular clouds, there is a tight relationship between the mean value of $R_{12/13}$ along a given line-of-sight and the $^{12}$CO column density along the same line-of-sight, and give a simple polynomial fitting formula for $R_{12/13}$ as a function of the $^{12}$CO column density. To produce our second set of $^{13}$CO maps, we therefore first calculate the $^{12}$CO column density along one axis of our datacubes, and then use these values to compute the mean value of $R_{12/13}$ for each line-of-sight using the \citet{SzucsEtAl2014} fitting formula. Finally, we derive the number density of $^{13}$CO in each cell along each line-of-sight by dividing the $^{12}$CO number density by the mean value of $R_{12/13}$. These $^{13}$CO number densities are then used to compute the $^{13}$CO emission in the same fashion as in our constant ratio models. We note that although this procedure neglects any spatial variations in $R_{12/13}$ along a given line-of-sight, \citet{SzucsEtAl2014} have shown that the resulting emission maps differ by only a few percent from those derived from a self-consistent (but costly) chemical treatment. The differences in the CVI statistics of the two sets of $^{13}$CO emission maps - those derived using a constant $R_{12/13}$ and those that use a spatially varying value of $R_{12/13}$ - will be discussed in Sec. \ref{subsec:ab13CO} below.

Although R{\sc admc}-3{\sc d} accounts for the dust continuum emission as well as the CO line emission, we make sure to subtract off the continuum term before analysing the statistical properties of the synthetic maps. For further information about the usage of R{\sc admc}-3{\sc d} and the calculation of the dust continuum and the line emission we refer the reader to \cite{Dullemond2012}.

The radiative transfer calculation yields position-position-velocity (PPV) cubes of brightness temperatures $T_B$, which are used to compute centroid velocities. We will refer to these as the ``intensity'' models. In analogy, we construct centroid velocity maps out of the PPP simulation data of the density and velocity field, as described in Sec. \ref{subsec:CVI}. We will refer to these as the ``density'' models.

\subsection{Centroid velocity increments}
\label{subsec:CVI}

Our study is based on the two-point statistics of line centroid velocities. Following \citet{LisEtAl1996}, we analyse the centroid $C(x, y) = C(\textbf{r})$ of line-of-sight projected velocities, defined as
\begin{equation}
\label{eq:CV}
C(\textbf{r}) = \frac{\int F(\textbf{r}, z) v_z(\textbf{r}, z)\text{d}z}{\int F(\textbf{r}, z)\text{d}z},
\end{equation}
where the variable $v_z(\textbf{r}, z)$ is the line-of-sight velocity in the z-direction. The quantity $F(\textbf{r}, z)$ is a statistical weight and denotes either the underlying density field or the brightness temperatures from the radiative transfer post-processing.

The centroid velocity increment $\delta C_{\ell}$ is defined as the separation between two points of the CV by spatial distances $\ell$, i.e.
\begin{equation}
\delta C_{\ell}(\textbf{r}) = C(\textbf{r}) - C(\textbf{r} + \textbf{\Bell}) .
\end{equation}
It thus connects centroid velocities of different regions in the plane.

There are many other techniques to measure velocity fluctuations as a function of spatial scale, e.g. the spectral correlation function \citep{RosolowskyEtAl1999}, the velocity channel analysis \citep{LazarianEtAl2000,LazarianEtAl2004}, the $\Delta$-variance method \citep{OssenkopfEtAl2008a,OssenkopfEtAl2008b} or the Principal Component Analysis \citep{HeyerAndSchloerb1997,BruntAndHeyer2002a}. The latter has been applied to study the gas dynamics of the same simulation data used in this paper in \citet{BertramEtAl2014}.

\subsection{Structure functions of the CVI}

\begin{figure}
\centerline{
\includegraphics[width=1.0\linewidth]{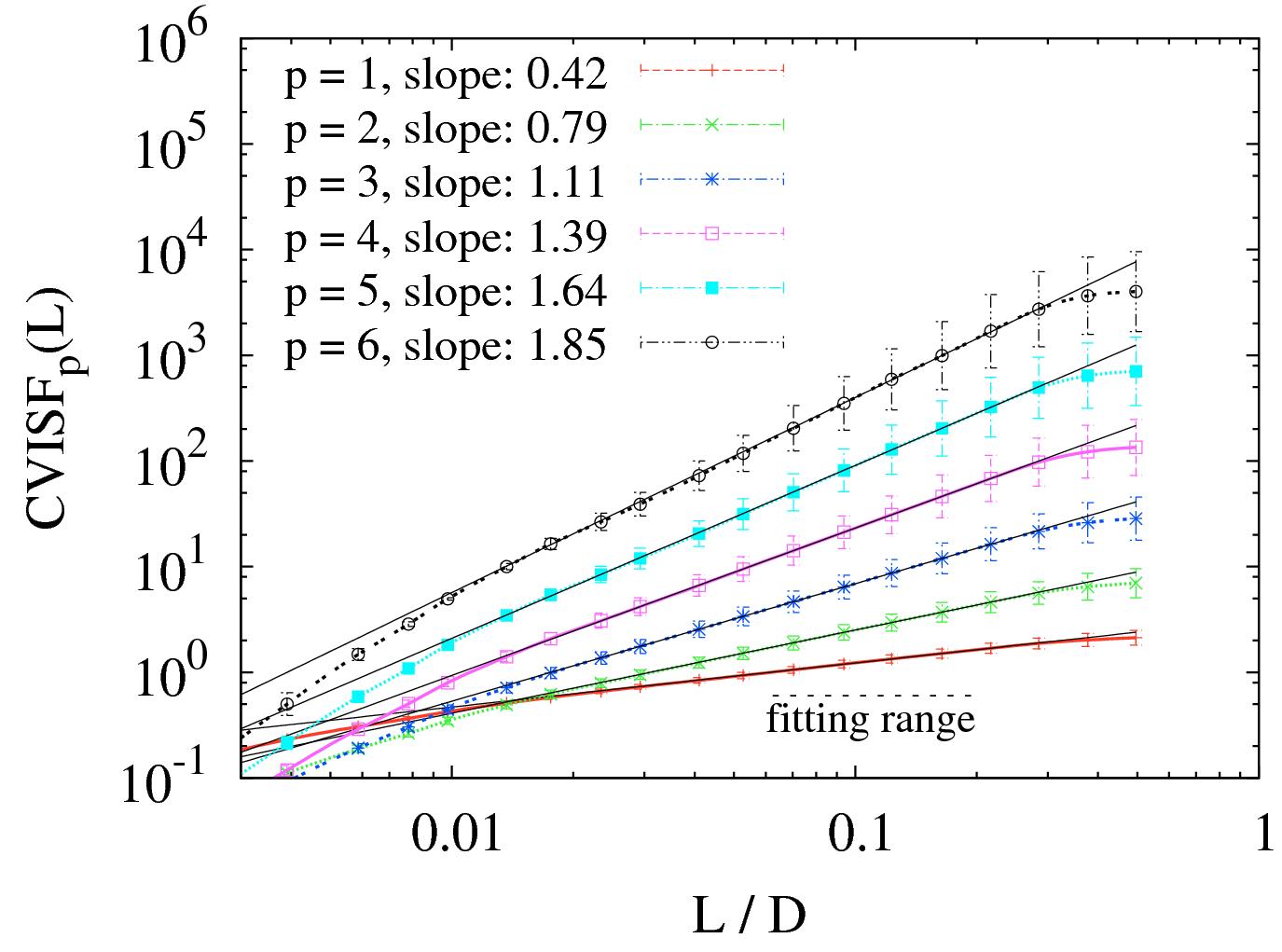} \\
}
\caption{CVI structure function for orders up to $p = 6$ with the inferred slopes $\zeta_p$ as a function of spatial scale for one snapshot in time, normalised by the total box size. The velocity here is weighted by the total density field for an initial number density of $n_{0} = 300$~cm$^{-3}$. The fitting range is indicated by a horizontal dashed line. Error bars denote 1-$\sigma$ variations for all 3 line-of-sight directions.}
\label{fig:SF}
\end{figure}

Statistical moments $p$ of the distribution of all possible CVI are called structure functions (CVISF). They are defined as
\begin{equation}
\text{CVISF}_p(\ell) = \langle |\delta C_{\ell}(\textbf{r})|^p \rangle
\end{equation}
where the average $\langle \rangle$ is computed over all possible CVI separated by $\ell$ in the plane. When $p$ increases, the CVISF give a stronger weight to rare events. In our study we have computed the CVISF for each possible line-of-sight direction x, y and z. Although we use a weak magnetic field in the z-direction, the turbulence in our simulations is trans-Alfv\'enic or mildly super-Alfv\'enic (see Table \ref{tab:setup}). The field lines are therefore dragged along with the turbulent flow, with the result that the turbulence remains approximately isotropic \citep[see, e.g.][]{BurkhartEtAl2014}. Hence, we do not expect significant variations of the CVISF along the different lines-of-sight. We therefore average all CVISF of the different line-of-sight directions and compute all CVISF for $p \le 6$. Finally, the CVISF were fit to power laws of the form
\begin{equation}
\text{CVISF}_p(\ell) \propto \ell^{\zeta_p} ,
\end{equation}
where $\zeta_p$ is a function that depends on the statistical moment $p$ of the CVI distribution. To calculate the scaling exponents, we use a fitting range from $1/16$ to $1/5$ of the total box size, as used by \citet{FederrathEtAl2010} and \citet{KonstandinEtAl2012a}. For a box with $D = 512$ grid cells for each side this translates to 32 and 102 cells. Figure \ref{fig:SF} shows an example of a CVISF for orders up to $p = 6$ for one snapshot in time, where the velocity is weighted by the total density field from model n300.

\subsection{Fourier and power spectra}
\label{subsec:powerspectra}

Fourier spectra are used to analyse the scale dependence of the kinetic energy of turbulence simulations. Assessing 3D turbulence power spectra, \citet{Kolmogorov1941} showed through statistical arguments that the kinetic energy should scale as $E(k) \propto k^{-2/3}$ in case of incompressible turbulence, where k is the wavenumber. Later, \citet{Burgers1948} derived $E(k) \propto k^{-1}$ for supersonic, compressive turbulence. We define the power spectrum as the change of the kinetic energy with its wavenumber, i.e. $P(k) = \text{d}E/\text{d}k$. The kinetic energy spectrum can be computed in Fourier space via
\begin{equation}
\label{eq:Ek}
E(k)\text{d}k = \frac{1}{2} \int \widehat{\bf{v}} \cdot \widehat{\bf{v}}^{*} 4\pi k^2 \text{d}k,
\end{equation}
where $\widehat{\bf{v}}$ denotes the Fourier transform of the velocity field and $\widehat{\bf{v}}^{*}$ its complex conjugate. Following \citet{Kolmogorov1941}, the kinetic energy down the cascade in the inertial range scales with $\langle \delta v_{\ell}^2 \rangle \propto \ell^{\zeta_2}$, where $\zeta_2$ is the slope of the second order structure function obtained in the inertial range (or fitting range, as we term it in our case). Thus, the power spectrum can be computed and we obtain
\begin{equation}
\label{eq:scaling}
P(k) = \frac{\text{d}E}{\text{d}k} \propto k^{-1-\zeta_2} = k^\alpha,
\end{equation}
where we have used $k \propto \ell^{-1}$ and set $\alpha = -1 - \zeta_2$ as the value of the spectral slope. Hence, the power spectrum can be determined by evaluating $\zeta_2$ of the second order structure function. Using Eq. \ref{eq:scaling}, we then obtain slopes $P(k) \propto k^{-5/3}$ and $P(k) \propto k^{-2}$ for both Kolmogorov ($\alpha = -5/3$) and Burgers ($\alpha = -2$) turbulence, respectively. In this paper, we compute power spectra for 2D maps of CV in order to obtain information about the underlying physics of the different chemical components.

\subsection{Theoretical relations for the SF slopes}
\label{subsec:theory}

Following \citet{Kolmogorov1941}, we assume that the kinetic energy scales as $E(k) \propto \langle \delta v^2 \rangle$ and $E(k) \propto k^{-2/3}$ in the case of incompressible turbulence, from which it directly follows that $\langle \delta v \rangle \propto \ell^{1/3}$. Furthermore, by taking $\langle \delta v_{\ell}^p \rangle \propto \ell^{\zeta_p}$, we can infer the theoretical scaling relation for the structure functions (SF) in the sense of Kolmogorov and find
\begin{equation}
\label{eq:K41}
\zeta_p = \frac{p}{3}.
\end{equation}
This law will be referred to as K41. Later on, \citet{SheAndLeveque1994} applied intermittency corrections to K41 and set $\zeta_p = p/3 + \tau_{p/3}$, since Eq. \ref{eq:K41} seems to deviate significantly from many experimental studies for $p > 3$ \citep{BenziEtAl1993,SheAndLeveque1994,Dubrulle1994,Frisch1995}. The function $\tau_{p/3}$ thereby includes additional terms that depend on $p$. \citet{SheAndLeveque1994} propose
\begin{equation}
\label{eq:SL94}
\zeta_p = \frac{p}{9} + 2\left[1 - \left(\frac{2}{3}\right)^{p/3}\right],
\end{equation}
which we refer to as the SL94 model. As shown by \citet{BoldyrevEtAl2002}, hereafter B02, and \citet{Dubrulle1994}, one important parameter in models of supersonic turbulence is the fractal co-dimension, $C = 3 - D$, where $D$ is the dimension of the most intermittent structures. The SL94 model assumes 1D filaments with $D = 1$ as the most intermittent structures, while the B02 model assumes sheet-like structures with $D = 2$. Hence, \citet{BoldyrevEtAl2002} and \citet{Dubrulle1994} generalised Eq. \ref{eq:SL94} and found
\begin{equation}
\label{eq:B02}
\zeta_p = \Theta (1 - \Delta)p + C (1 - \Sigma^{\Theta p}),
\end{equation}
where $\Sigma = 1 - \Theta / C$. The other parameters $\Theta$ and $\Delta$ enter the theory through the so-called Kolmogorov refined similarity hypothesis, relating the scaling of fluctuations of the velocity field to the scaling of the fluctuating energy dissipation. We leave those two as free arbitrary parameters for now.

\section{Results}
\label{sec:results}

\begin{table}
\begin{tabular}{l|c|c|c}
\hline\hline
Model name & n30 & n100 & n300 \\
\hline
Mean density $\lbrack$cm$^{-3}\rbrack$ & 30 & 100 & 300 \\
Resolution & 512$^3$ & 512$^3$ & 512$^3$ \\
Box size [pc] & 20 & 20 & 20 \\
$t_{\text{end}}$ [Myr] & 5.7 & 5.7 & 5.7 \\
$\langle \Ma_{\text{s}} \rangle$ & 5.1 & 6.8 & 10.6 \\
$\langle \Ma_{\text{A}} \rangle$ & 1.0 & 1.1 & 1.5 \\
$\sigma_{\rho} / \langle \rho \rangle$ & 4.6 & 3.0 & 3.0 \\
$\langle x_{\text{H}_2} \rangle_{\text{mass}}$ & 0.61 & 0.78 & 0.98 \\
$\langle n_{\text{H}_2} \rangle_{\text{vol}}$ $\lbrack$cm$^{-3}\rbrack$ & 8.6 & 36.8 & 139.3 \\
$\langle n_{\text{H}_2} \rangle_{\text{mass}}$ $\lbrack$cm$^{-3}\rbrack$ & 263.6 & 447.1 & 1455.8 \\
$\langle n_{\text{CO}} \rangle_{\text{vol}}$ $\lbrack$cm$^{-3}\rbrack$ & $1.4 \times 10^{-4}$ & $1.7 \times 10^{-3}$ & $2.2 \times 10^{-2}$ \\
$\langle n_{\text{CO}} \rangle_{\text{mass}}$ $\lbrack$cm$^{-3}\rbrack$ & $2.3 \times 10^{-2}$ & $4.3 \times 10^{-2}$ & 0.3 \\
$\langle T \rangle_{\text{vol}}$ $\lbrack$K$\rbrack$ & 223.0 & 68.4 & 34.9 \\
$\langle T \rangle_{\text{mass}}$ $\lbrack$K$\rbrack$ & 56.7 & 25.8 & 12.7 \\
$\langle N_{\text{H}_2} \rangle$ $\lbrack$cm$^{-2}\rbrack$ & $5.3 \times 10^{20}$ & $2.3 \times 10^{21}$ & $8.6 \times 10^{21}$ \\
$\langle N_{\text{CO}} \rangle$ $\lbrack$cm$^{-2}\rbrack$ & $8.7 \times 10^{15}$ & $1.1 \times 10^{17}$ & $1.3 \times 10^{18}$ \\
\hline
\end{tabular}
\caption{Overview of the different models with some characteristic values for the last snapshot. From top to bottom we list: mean number density, resolution, box size, time of the last snapshot, mean sonic Mach number, mean Alfv\'en Mach number, ratio of density dispersion and mean density, mass-weighted mean abundances of H$_2$ \citep[i.e. the percentage of atomic hydrogen that has been converted to H$_2$, see][]{GloverAndMacLow2011}, mean volume- and mass-weighted H$_2$ and CO number density, mean volume- and mass-weighted temperature and mean column density of H$_2$ and CO.}
\label{tab:setup}
\end{table}

\begin{figure*}
\centerline{
\includegraphics[height=0.27\linewidth]{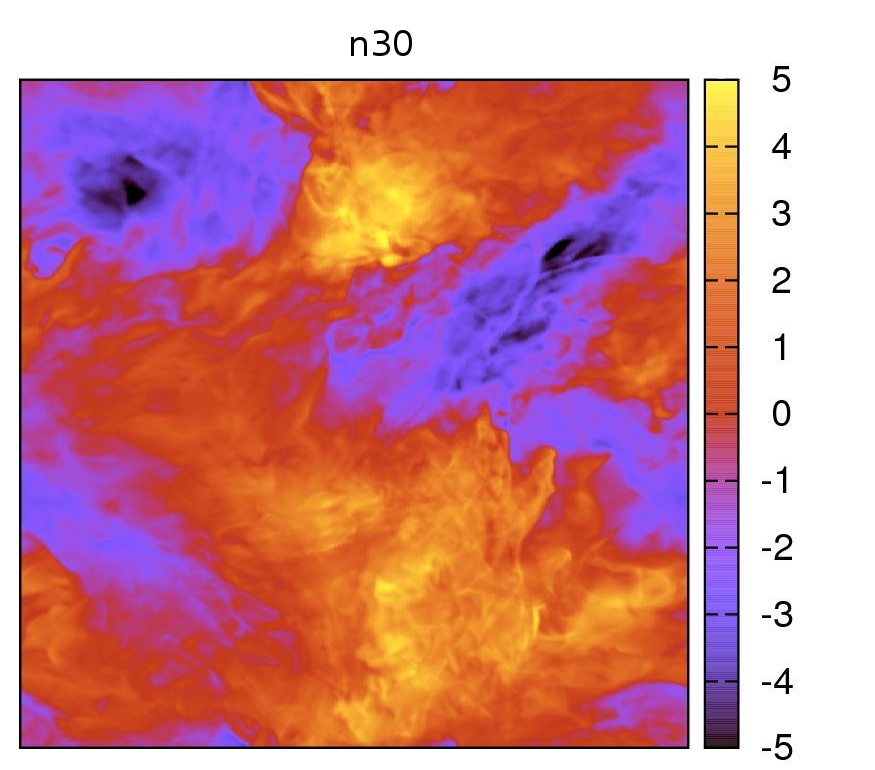}
\includegraphics[height=0.27\linewidth]{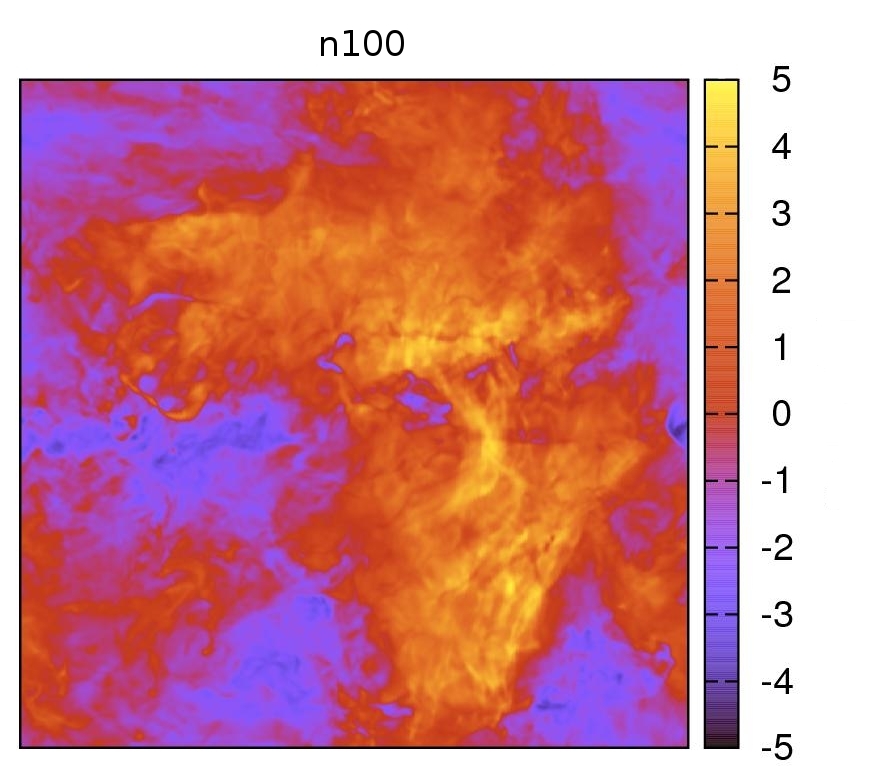}
\includegraphics[height=0.27\linewidth]{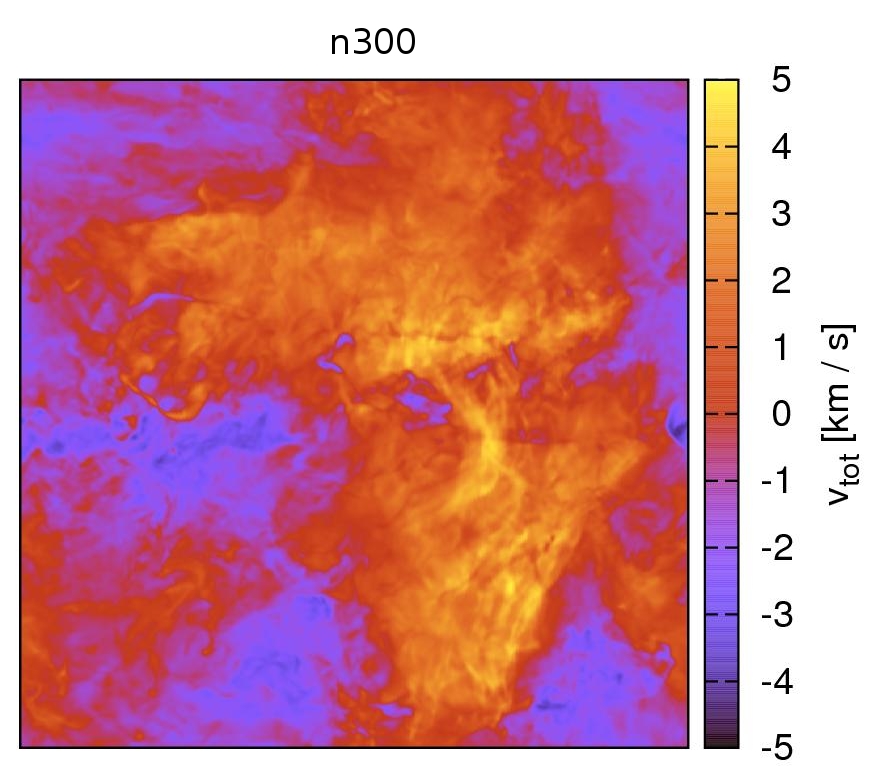}
}
\centerline{
\includegraphics[height=0.25\linewidth]{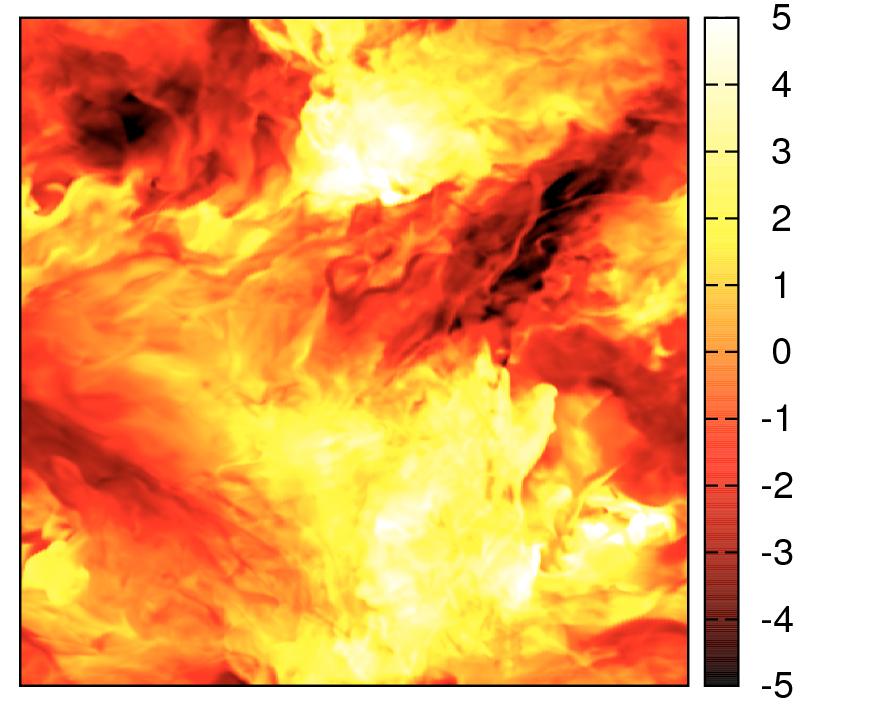}
\includegraphics[height=0.25\linewidth]{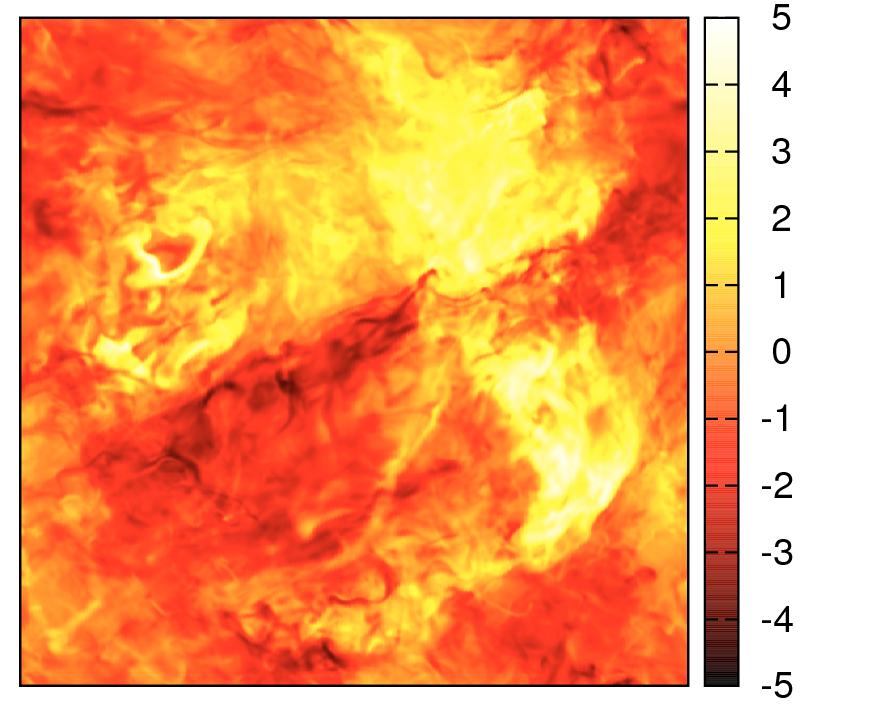}
\includegraphics[height=0.25\linewidth]{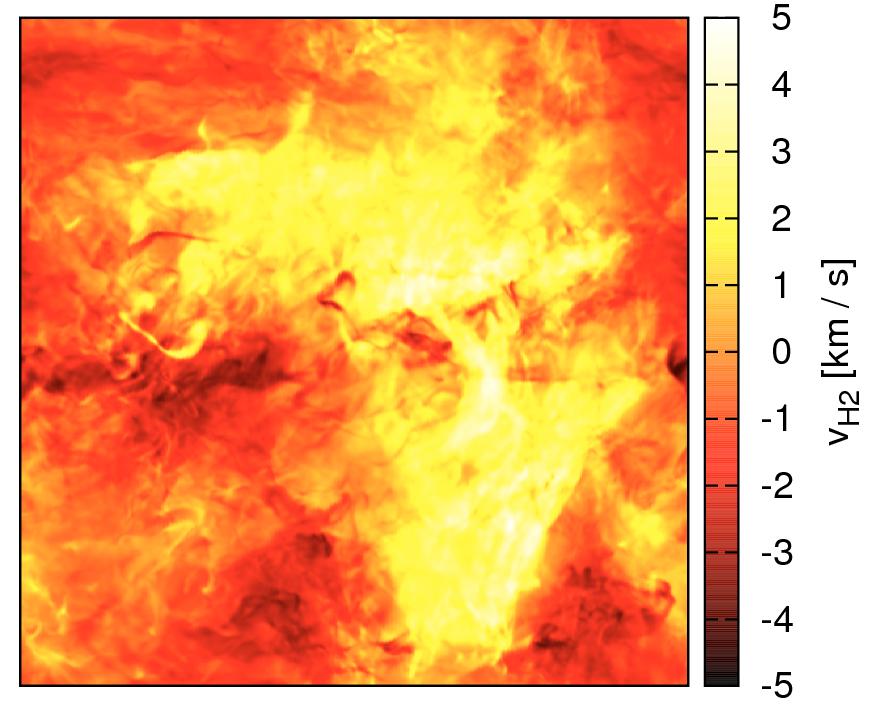}
}
\centerline{
\includegraphics[height=0.25\linewidth]{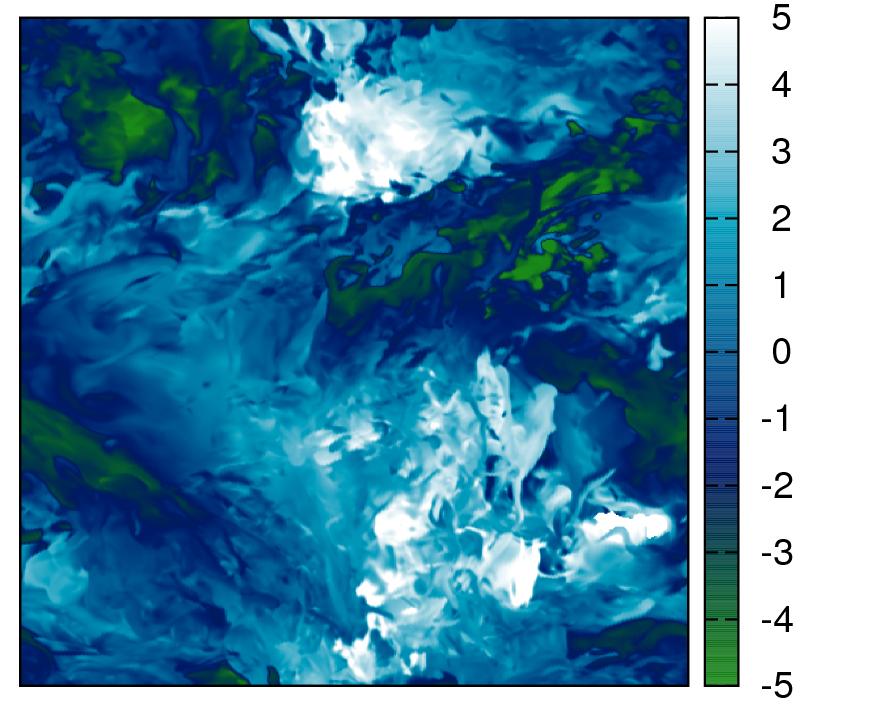}
\includegraphics[height=0.25\linewidth]{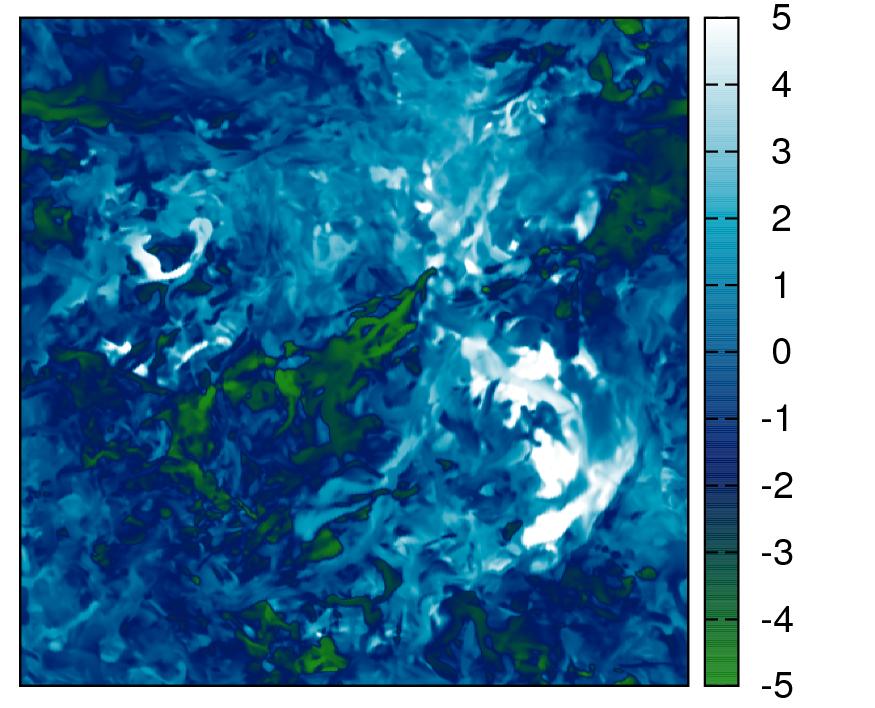}
\includegraphics[height=0.25\linewidth]{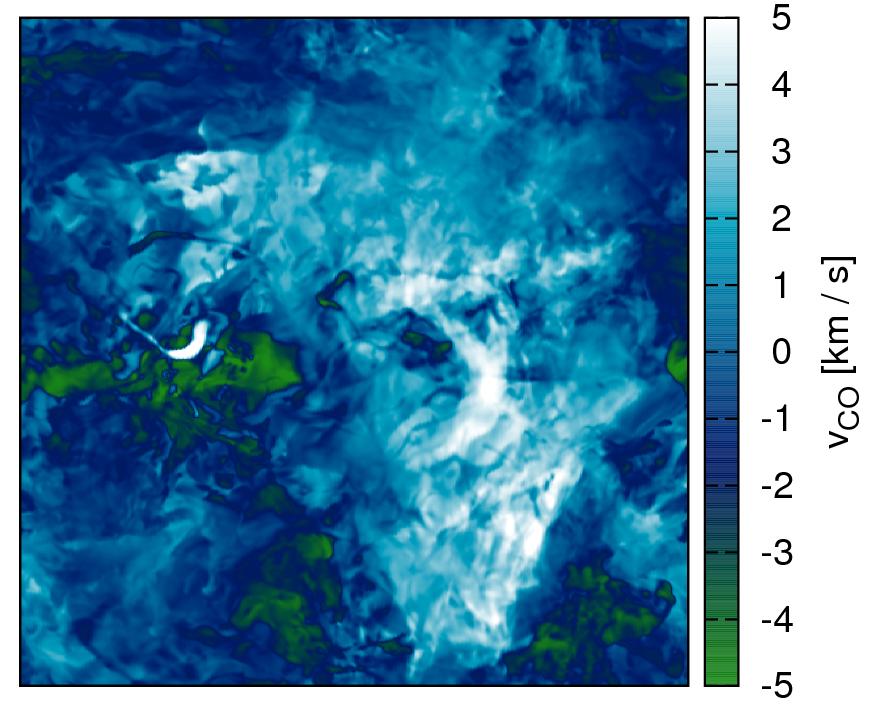}
}
\centerline{
\includegraphics[height=0.25\linewidth]{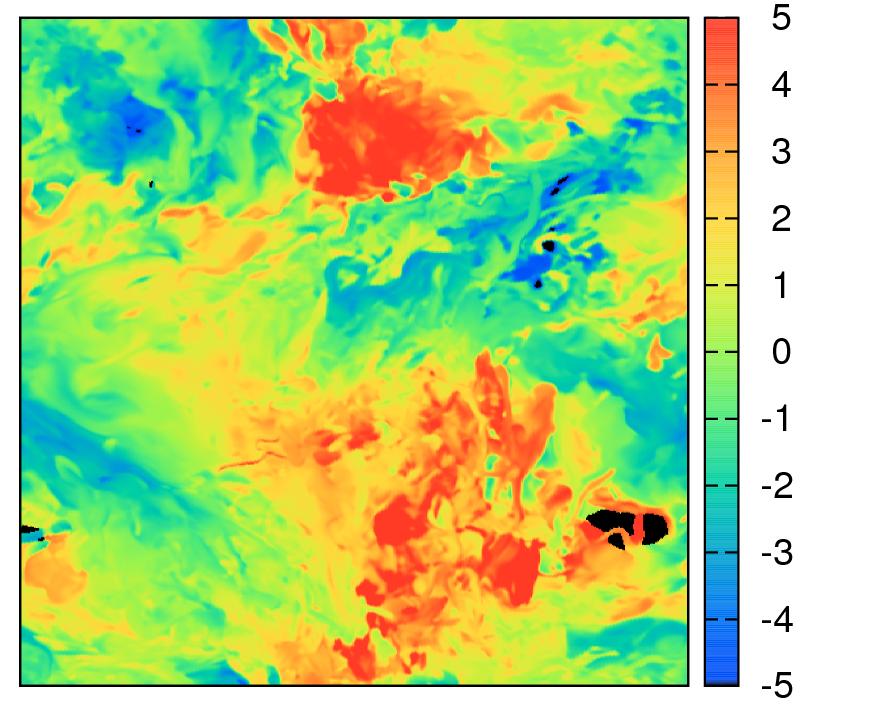}
\includegraphics[height=0.25\linewidth]{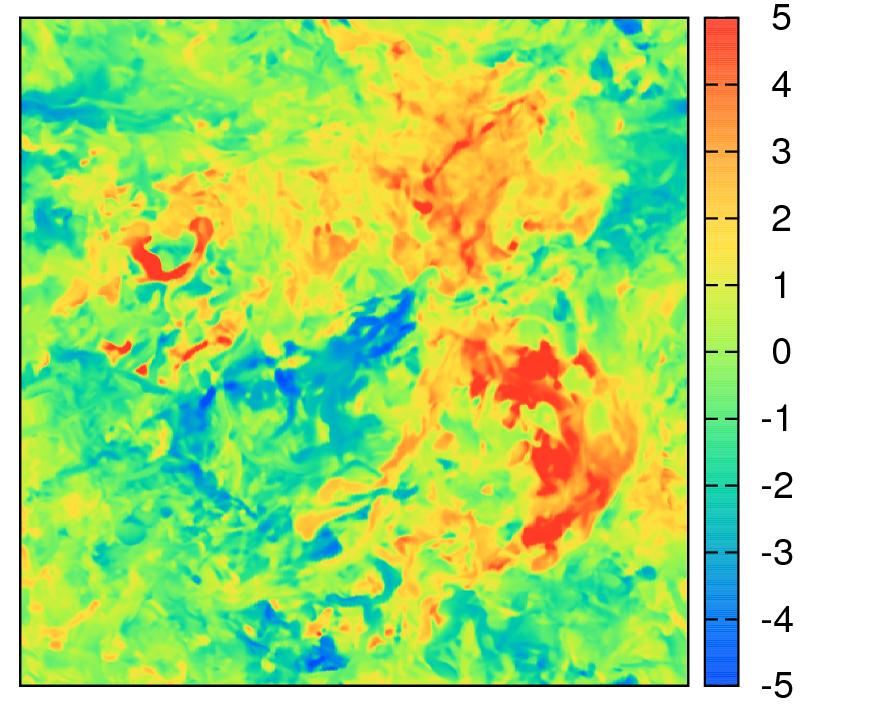}
\includegraphics[height=0.25\linewidth]{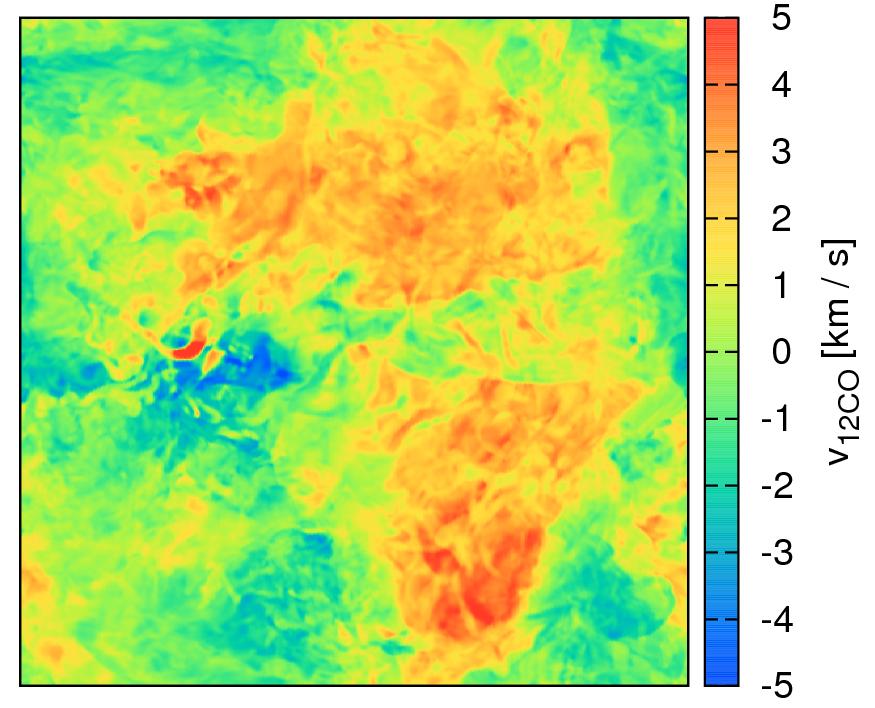}
}
\centerline{
\includegraphics[height=0.25\linewidth]{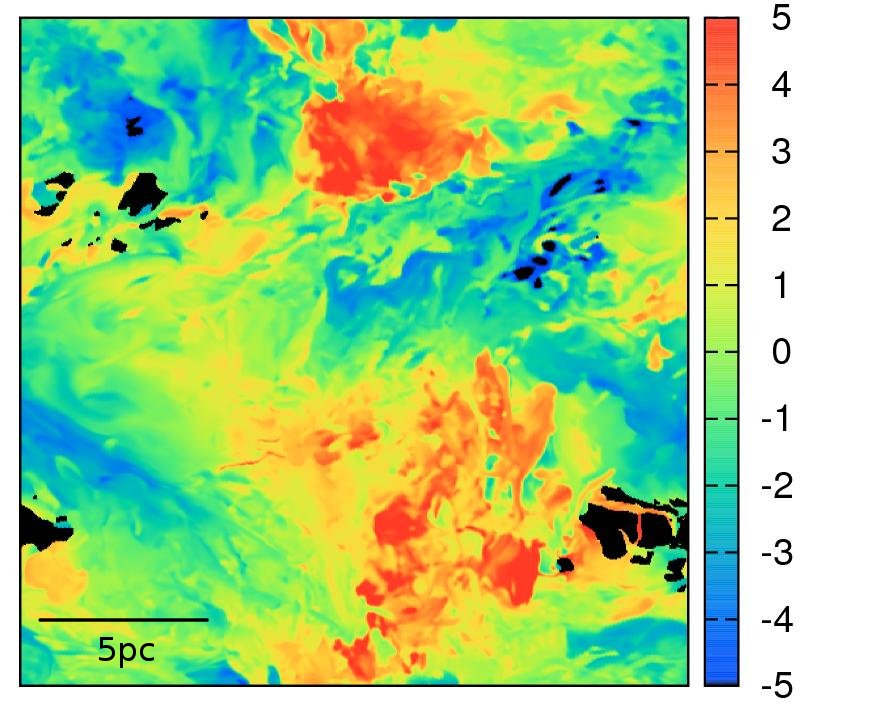}
\includegraphics[height=0.25\linewidth]{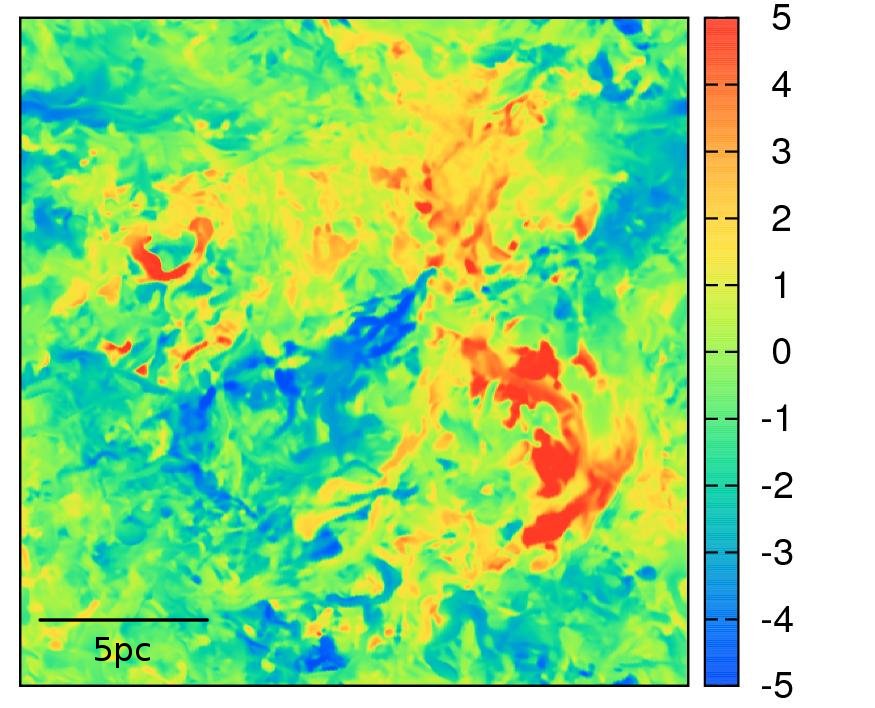}
\includegraphics[height=0.25\linewidth]{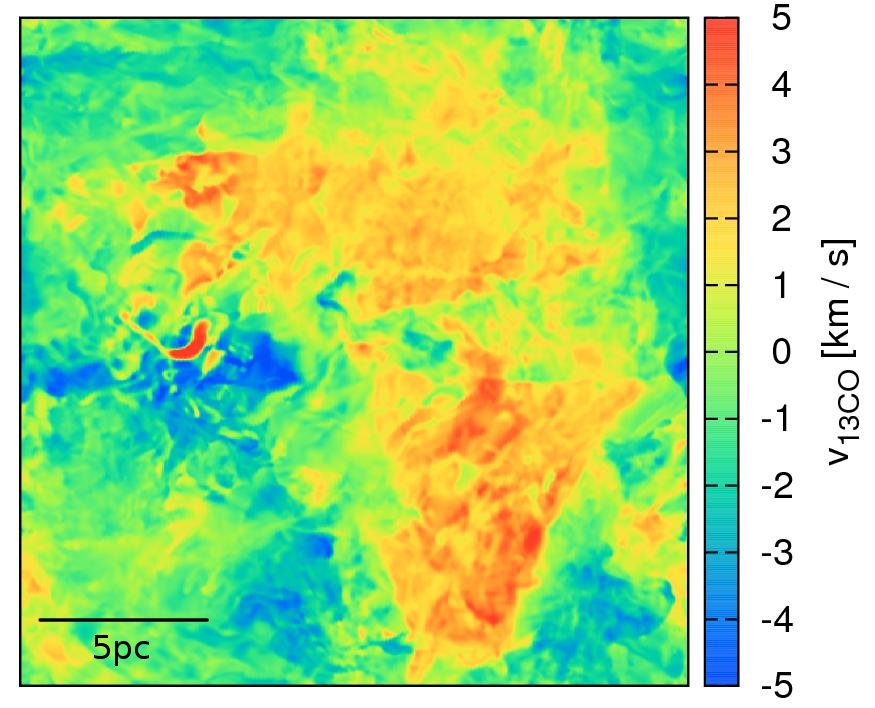}
}
\caption{Centroid velocities in z-direction for models n30 (left column), n100 (middle column) and n300 (right column). From top to bottom: CV maps of the total density, H$_2$ and $^{12}$CO density followed by CV maps of the $^{12}$CO and $^{13}$CO integrated intensities. Each side has a length of $20\,$pc. Note that the velocity field of the n30 model uses a different turbulent seed than the n100 and the n300 model. Black areas in the $^{12}$CO and $^{13}$CO map of the n30 model denote regions where the brightness temperatures are zero along the LoS.}
\label{fig:images}
\end{figure*}

We analyse different numerical models with and without radiative transfer post-processing and vary the initial number density in the simulation box, in order to study the influence on the structure functions. Table~\ref{tab:setup} gives an overview of the numerical models that we study in this paper. We quote both mass- and volume-weighted quantities, which we define using the expressions $\langle f \rangle_{\text{mass}} = \sum f \rho dV / \sum \rho dV$ and $\langle f \rangle_{\text{vol}} = \sum f dV / \sum dV$, respectively. More detailed analyses of the H$_{2}$ and CO distributions produced in this kind of turbulent simulation can be found in previous studies by \citet{GloverEtAl2010} and \citeauthor{ShettyEtAl2011a}~(2011a). Examples of the integrated intensity and column density PDFs are shown in \citeauthor{ShettyEtAl2011a}~(2011a). Figure \ref{fig:images} shows the centroid velocity maps computed via Eq. \ref{eq:CV} for all models and for all chemical components.

\subsection{Analysis of the CVISF}
\label{subsec:analysisCVISF}

\begin{figure}
\centerline{
\includegraphics[height=0.7\linewidth,width=1.0\linewidth]{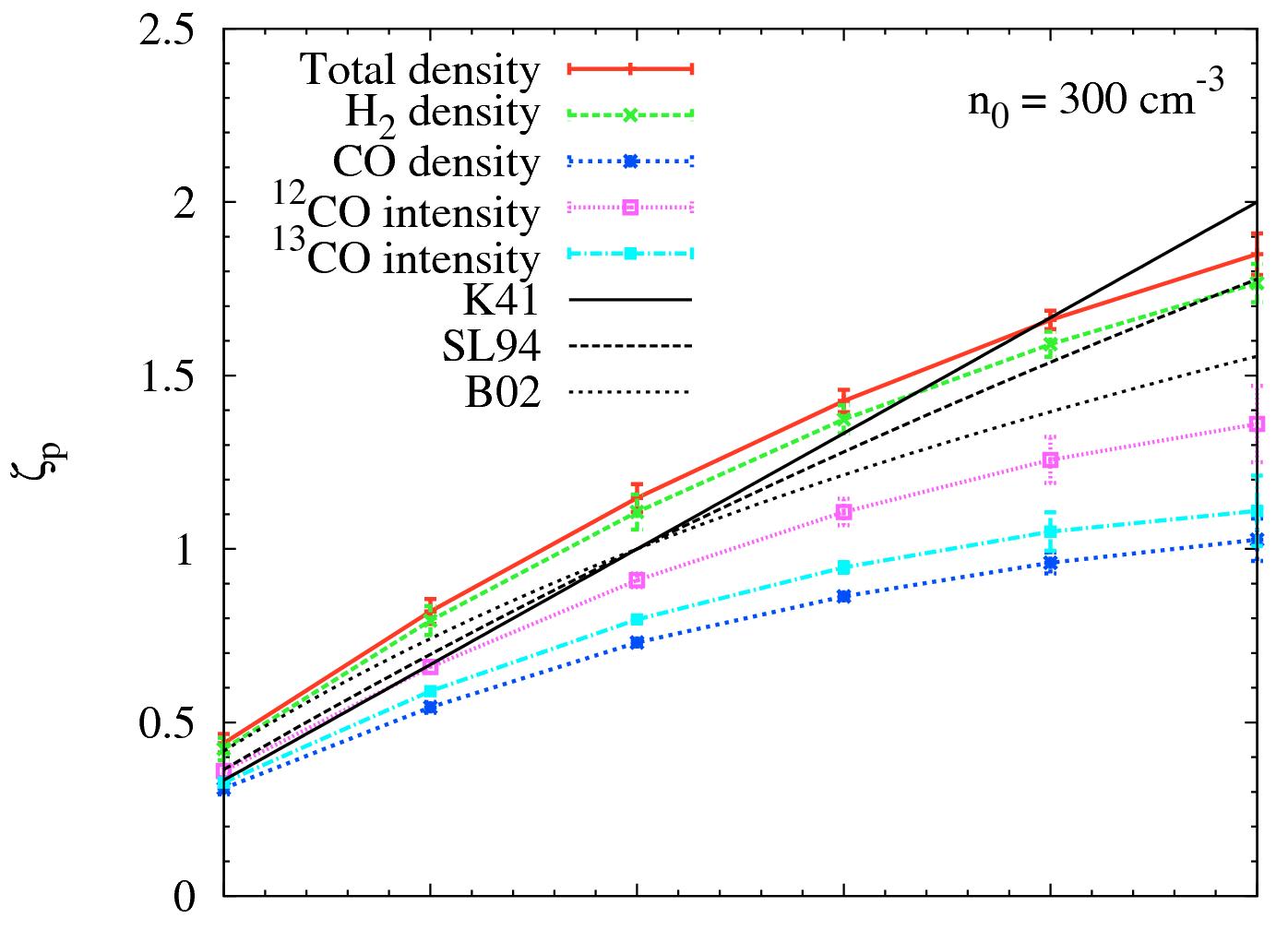} \\
}
\centerline{
\includegraphics[height=0.7\linewidth,width=1.0\linewidth]{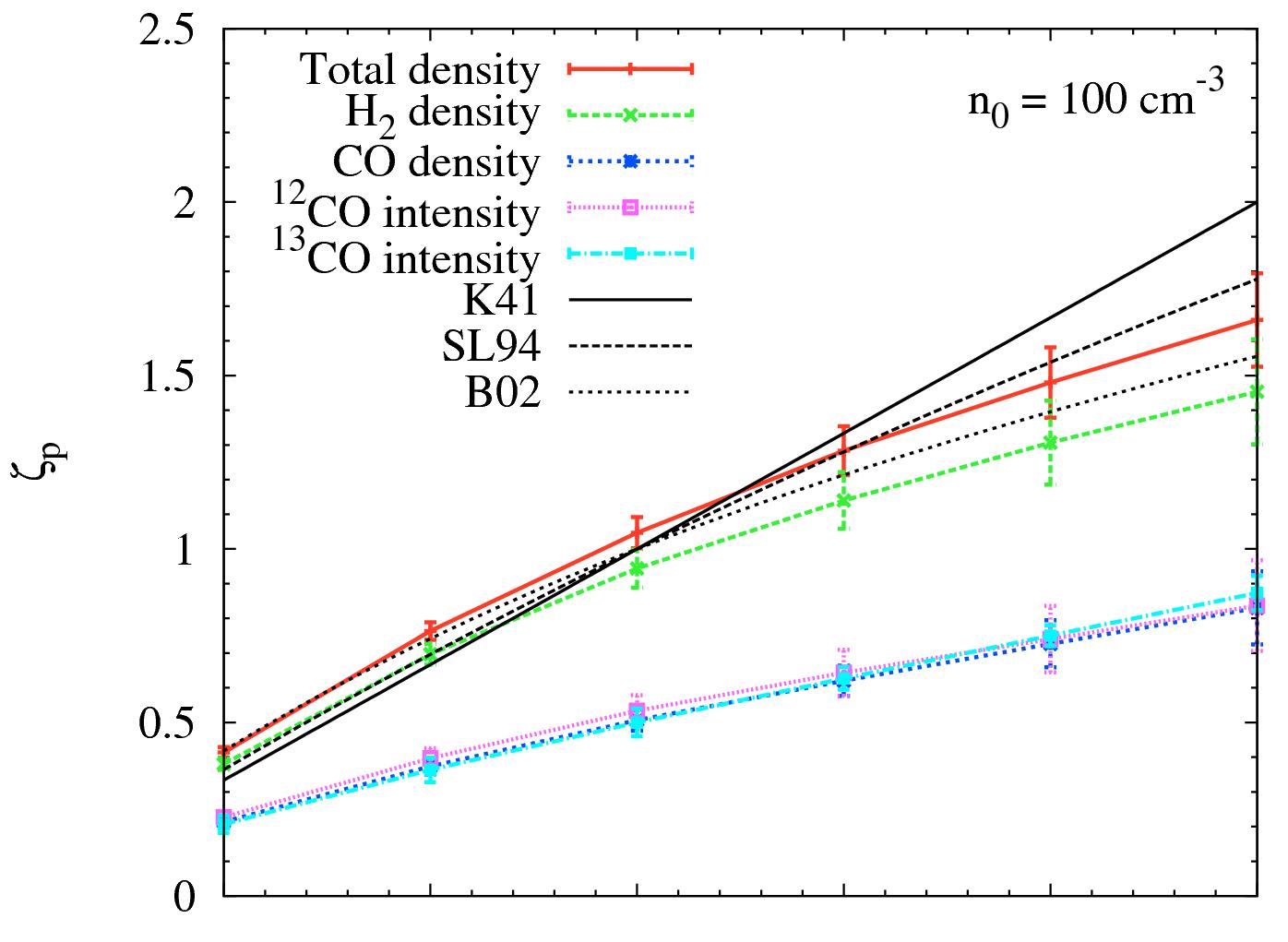} \\
}
\centerline{
\includegraphics[height=0.75\linewidth,width=1.0\linewidth]{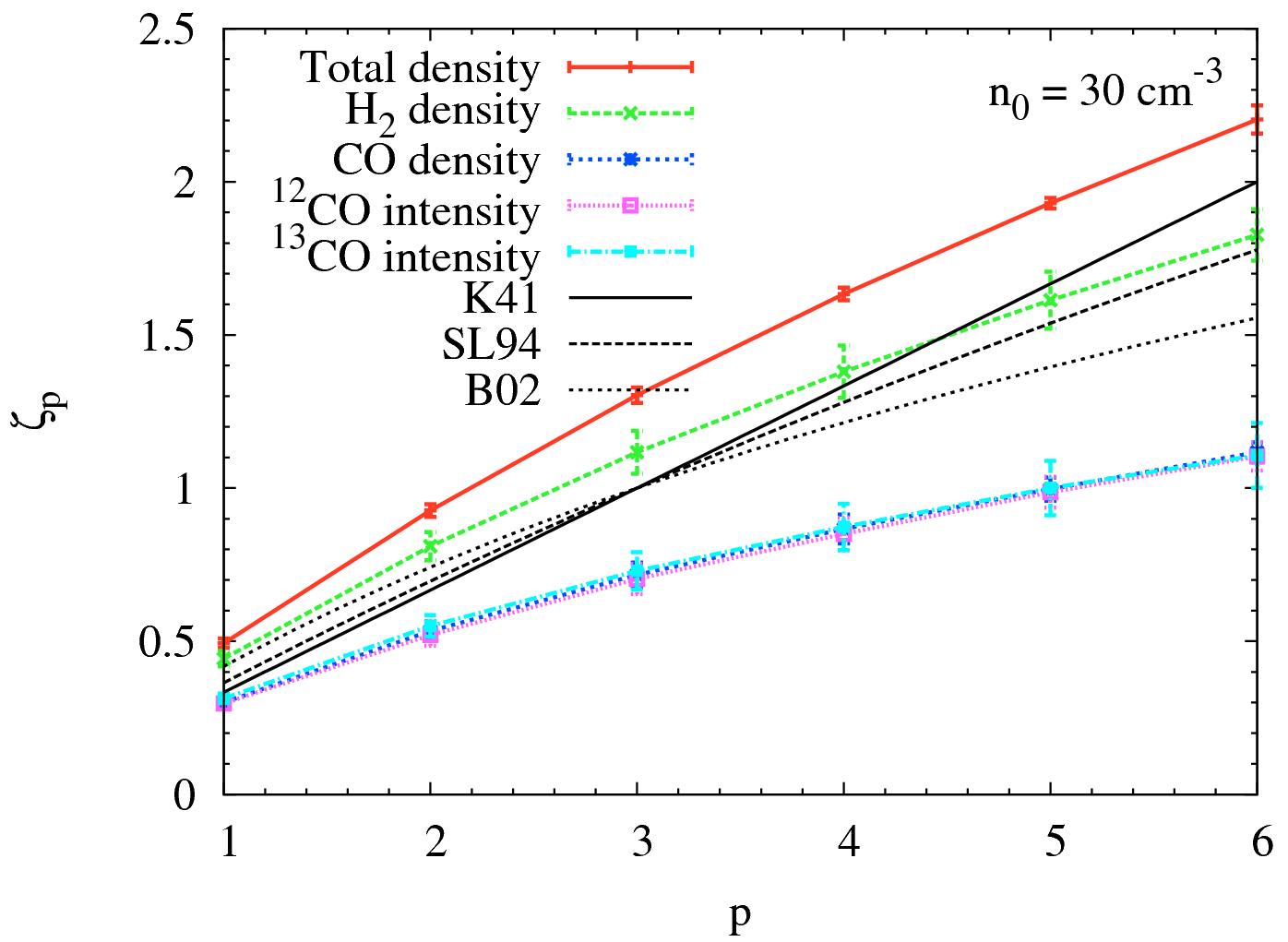} \\
}
\caption{Slopes $\zeta_p$ of the structure functions against order $p$ for all chemical models, i.e. the total density, H$_2$ density and the CO density model as well as the $^{12}$CO and $^{13}$CO intensity model. For comparison, we also show the theoretical scaling relations (black lines) from \citet{Kolmogorov1941}, \citet{SheAndLeveque1994} and \citet{BoldyrevEtAl2002}, denoted as K41, SL94 and B02. From top to bottom: runs of the different initial number densities, i.e. $n_{0} = 300, 100$ and $30 \: {\rm cm^{-3}}$. Error bars denote temporal and spatial 1-$\sigma$ fluctuations. Please note that a direct comparison of the theoretical models with our CV statistics is complicated due to the projection along the LoS. For a more detailed discussion about the influence of the projection, we refer the reader to Sec. \ref{subsec:projection} and \ref{subsec:fractaldim}.}
\label{fig:CVISF}
\end{figure}

We compute the CVISF for all models (n30, n100, n300) and all chemical tracers: the total density, H$_2$ density, CO density, $^{12}$CO and $^{13}$CO intensity. We average slopes of the CVISF using 3 snapshots in time (with 3 line-of-sight directions each) where we can assume both the chemistry and the turbulence to be in a stationary and converged state. In Figure \ref{fig:CVISF} we show slopes $\zeta_p$ of the CVISF as a function of the order $p$ for all models and chemical tracers. The error bars denote temporal and spatial 1-$\sigma$ fluctuations from the different time snapshots. Moreover, we also show the K41, SL94 and B02 scaling relations in the plots. We have to keep in mind that a direct comparison of the theoretical models with our CV statistics is nontrivial and sometimes impossible, since we analyse velocity centroids, while the theory predicts scaling relations for the full turbulent velocity field. For a more detailed discussion about the influence of the projection, we refer the reader to Sec. \ref{subsec:projection} and \ref{subsec:fractaldim}.

We find differences between the various models and tracers. In our n300 model we obtain similar slopes of the total density and the H$_2$ density model as well as for the CO density and the $^{13}$CO intensity model within their error bars. The slopes of the total density and the H$_2$ density model are steeper compared to the different CO tracer models, which flatten to higher order $p$ of the CVISF. The slopes $\zeta_p$ of the $^{12}$CO intensity model make up an intermediate case. They are slightly steeper than the CO density and $^{13}$CO intensity model, but flatter than those of the total density and H$_2$ density model.

Regarding the n100 model, we find consistent values for $\zeta_p$ for all CO tracer cases and for all orders $p$. Furthermore, we again measure steeper slopes for the total density and H$_2$ density model, which now slightly deviate from each other at higher order of the CVISF.

In our n30 model we obtain a larger discrepancy between the total density and the H$_2$ density model. Although the temporal and spatial 1-$\sigma$ fluctuations are large for both cases, we find a clear separation of the slopes for higher $p$. Again, all different CO tracers agree with each other within their error bars and are significantly flatter and those of the H$_2$ and total density cases.

\subsection{Analysis of the Fourier spectra}
\label{subsec:analysisPS}

\begin{figure}
\centerline{
\includegraphics[height=0.7\linewidth,width=1.0\linewidth]{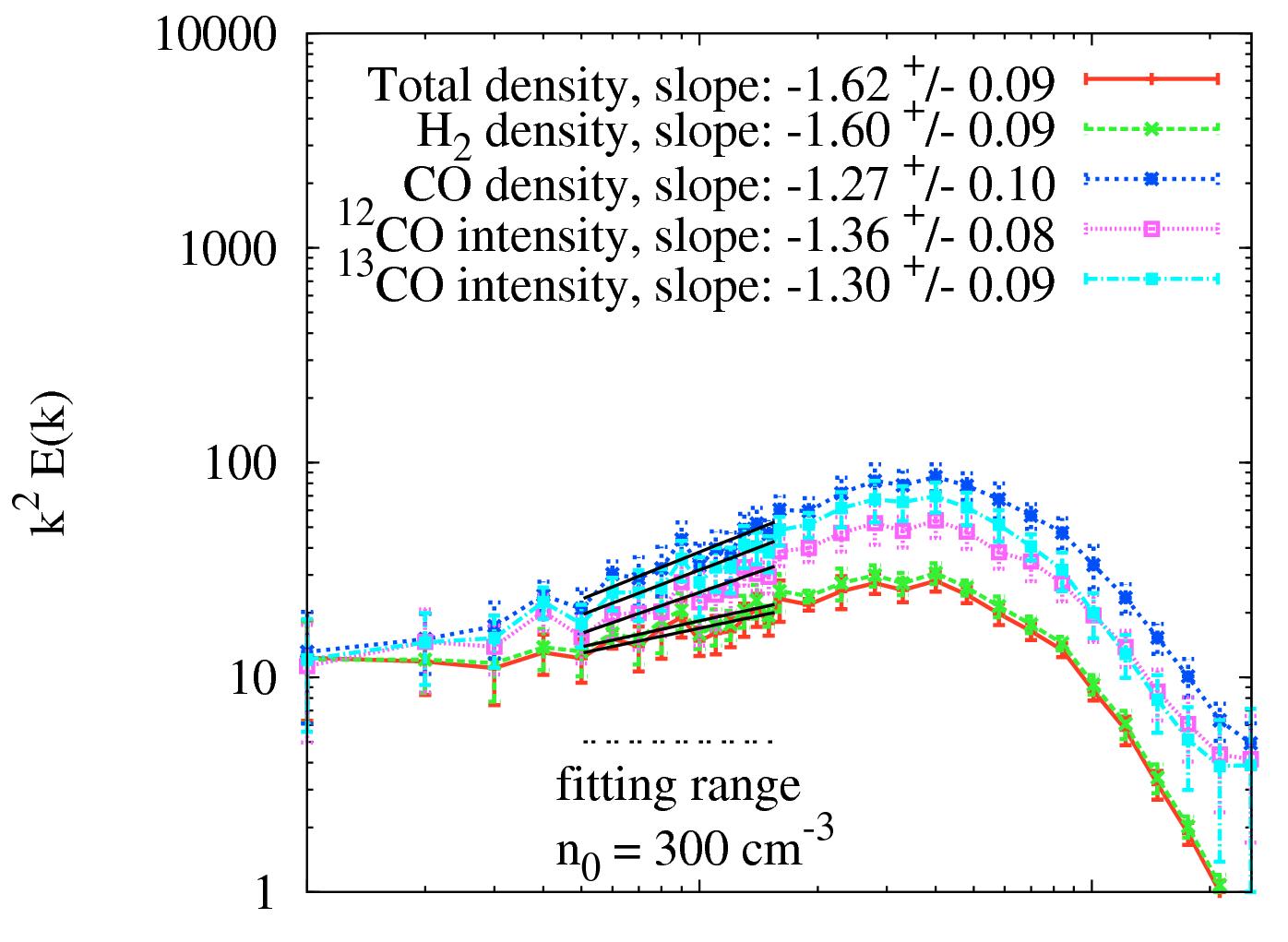} \\
}
\centerline{
\includegraphics[height=0.7\linewidth,width=1.0\linewidth]{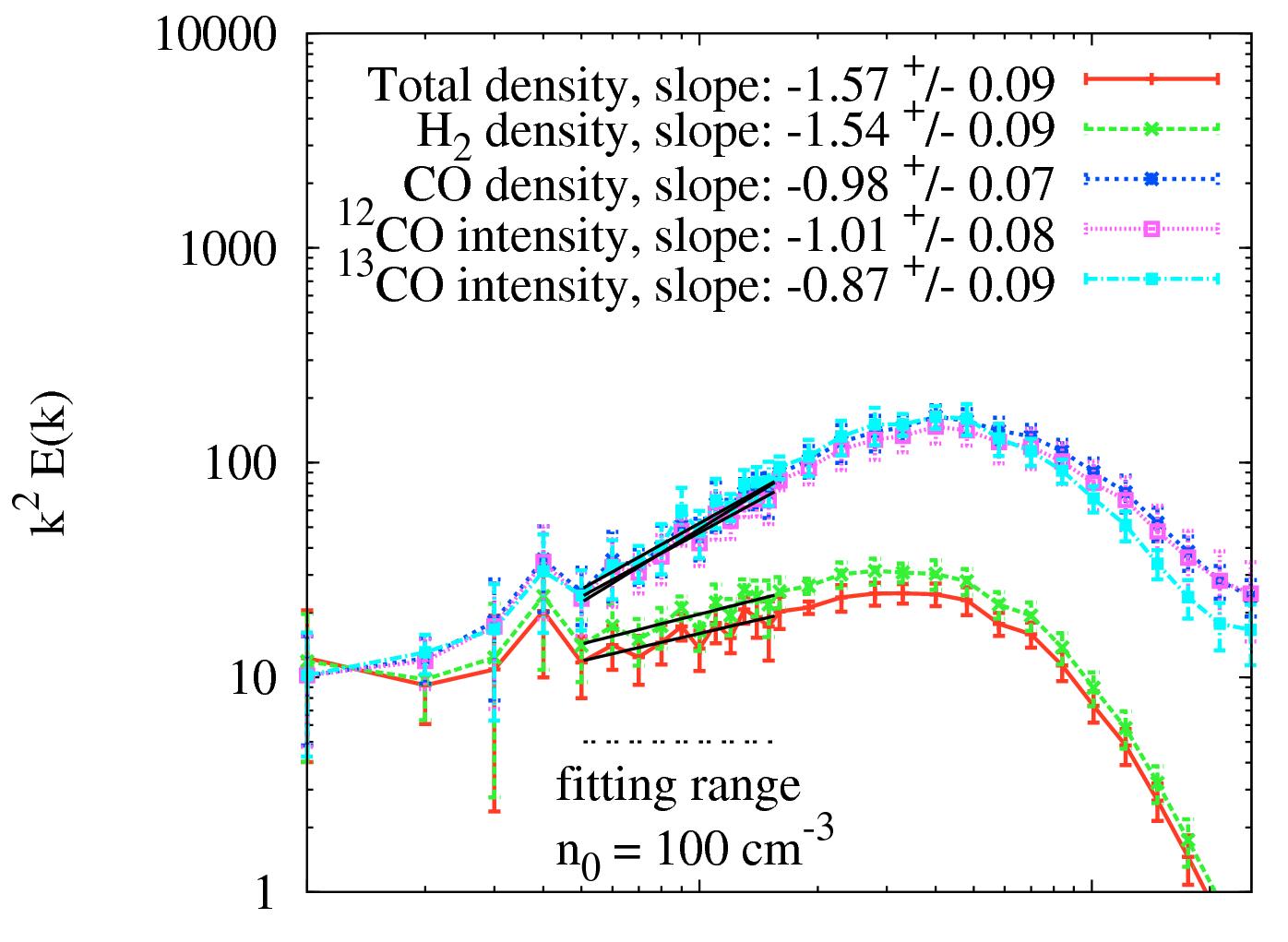} \\
}
\centerline{
\includegraphics[height=0.75\linewidth,width=1.0\linewidth]{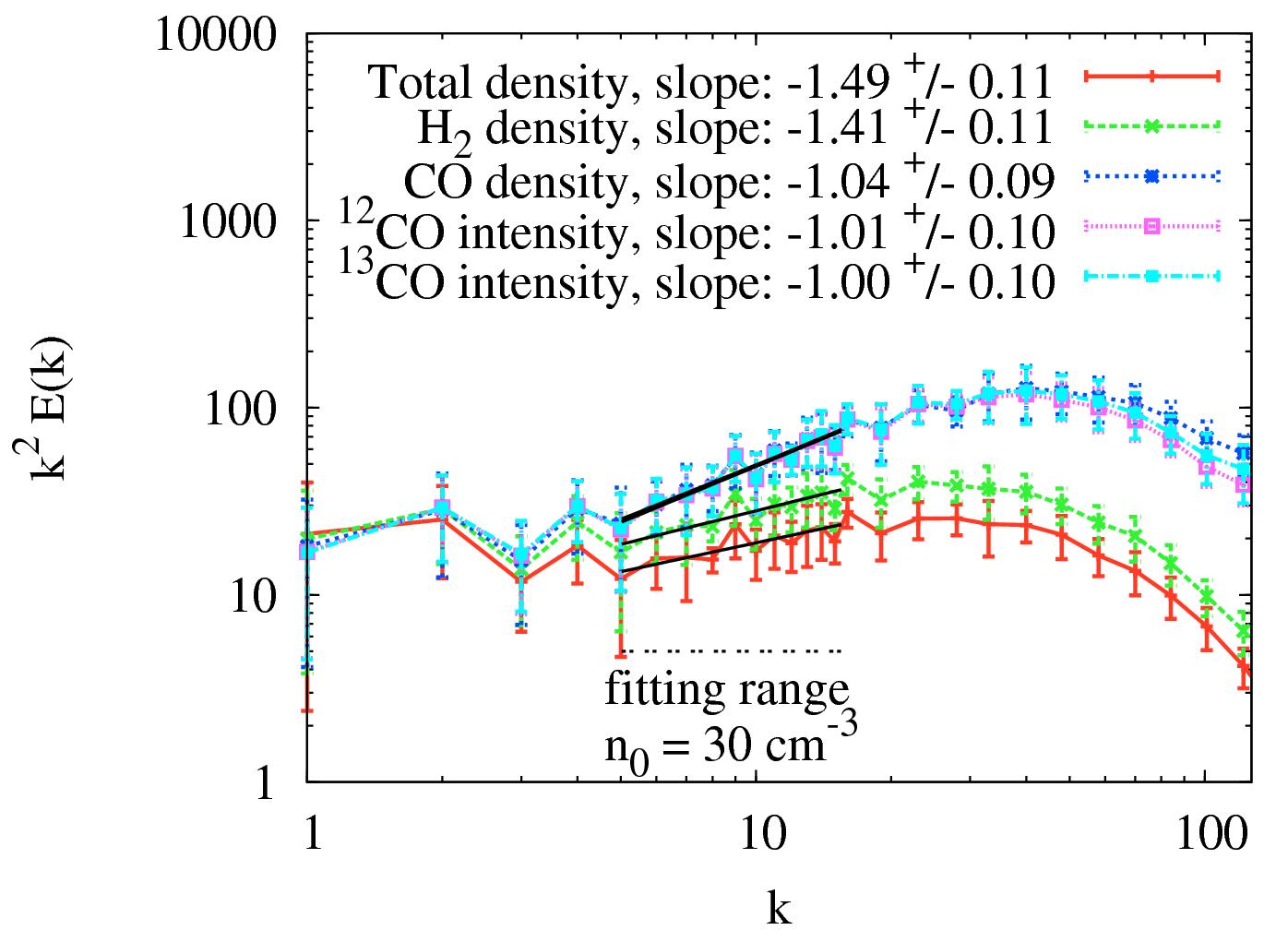} \\
}
\caption{Fourier energy spectra multiplied by $k^2$ as a function of scale $k$ for all density models and chemical components, i.e. the total density, H$_2$ density and the CO density model as well as the $^{12}$CO and $^{13}$CO intensities. From top to bottom: runs of the different initial number densities, i.e. $n_{0} = 300, 100$ and $30 \: {\rm cm^{-3}}$. Error bars denote temporal and spatial 1-$\sigma$ fluctuations. In all models, the total density and H$_2$ density cases show a significantly steeper slope compared to the CO tracer components. The fitting range is indicated by a horizontal dashed line. Slopes with the fitting errors are given in each plot for the different species.}
\label{fig:Spectra}
\end{figure}

Following Eq. \ref{eq:Ek}, we compute Fourier spectra and the scaling exponents of the power spectra by fitting a linear function in log-log space to the Fourier spectra in the fitting range of our simulations. The fitting range in $k$-space translates to wavenumbers from 5 to 16 and has to be chosen carefully due to the artificial accumulation of energy further down the cascade, known as the bottleneck effect \citep{Falkovich1994,DoblerEtAl2003,KanedaEtAl2003,HaugenAndBrandenburg2004,Kritsuk2007,BeresnyakAndLazarian2009}. In analogy to Figure \ref{fig:CVISF}, Figure \ref{fig:Spectra} shows with $k^2$ compensated Fourier spectra for the different density models and all chemical components.

In the n300 model we again find a good correlation between the total density and the H$_2$ density. We measure slopes $\alpha$ of $-1.62 \pm 0.09$ for the former and $-1.60 \pm 0.09$ for the latter, which agree well within their errors. For the CO density, $^{12}$CO and $^{13}$CO intensity cases we find slopes of $-1.27 \pm 0.10$, $-1.36 \pm 0.08$ and $-1.30 \pm 0.09$, which do not show significant differences within their errors.

The same trends are seen in the other density models. In the n100 model we again do not find a significant difference between the total density and the H$_2$ density, for which we measure slopes $\alpha$ of $-1.57 \pm 0.09$ and $-1.54 \pm 0.09$, respectively. For the CO density, $^{12}$CO and $^{13}$CO intensity cases we find very flat slopes of $-0.98 \pm 0.07$, $-1.01 \pm 0.08$ and $-0.87 \pm 0.09$.

In the low-density n30 model, the slopes $\alpha$ of the total density ($-1.49 \pm 0.11$) and H$_2$ density ($-1.41 \pm 0.11$) cases agree well within their errors, while the slopes of the CO density, $^{12}$CO and $^{13}$CO intensity cases are again much flatter and we measure $-1.04 \pm 0.09$, $-1.01 \pm 0.10$ and $-1.00 \pm 0.10$.

Overall, we find a significant difference in the slopes of the H$_2$ density and all CO tracers for all three density models. Slopes of the total density and H$_2$ density are found to be significantly steeper than those produced by the different CO tracers, while we do not obtain significant differences in the slopes between the different CO tracers within their errors.

\subsection{Variation of the abundance of $^{13}$CO}
\label{subsec:ab13CO}

Our numerical simulations follow the chemistry of $^{12}$CO, but not of its isotope $^{13}$CO. Consequently, in order to generate the data on the number density of $^{13}$CO that we need for our radiative transfer calculation, it is necessary to specify a conversion factor between $^{12}$CO and $^{13}$CO. As we explain in Sec. \ref{subsec:RADMC}, in most of our analysis, we assume a fixed $^{12}$CO to $^{13}$CO ratio, $R_{12/13} = 50$. However, in real molecular clouds, we expect spatial variations in $R_{12/13}$ in any regions in which the carbon is not all locked up in CO, caused by the two effects of chemical fractionation and selective photodissociation \citep{RoelligAndOssenkopf2013,SzucsEtAl2014}. To explore the effect that these variations may have on the CVI statistics of the $^{13}$CO emission maps, we produce a set of $^{13}$CO emission maps, using a spatially varying $R_{12/13}$ generated following the prescription given in Sec. \ref{subsec:RADMC} and in \citet{SzucsEtAl2014}. After applying the radiative transfer post-processing to both models, we compute the two maps of the CV and compare the slopes of the CVISF.

Although we find slight differences in the CV maps on small scales, the slopes $\zeta_p$ of the CVISF are essentially the same in both cases. The differences between the results in the two models are smaller than the 1-$\sigma$ temporal and spatial errors in the different snapshots. We thus conclude that the use of a constant $^{12}$CO/$^{13}$CO ratio is sufficient for obtaining proper slopes of the CVISF for $^{13}$CO.

\section{Discussion}
\label{sec:discussion}

\subsection{Slopes of the CVISF}
\label{subsec:slopesCVISF}

As shown in Figure \ref{fig:CVISF}, we find large differences in the CVISF slopes for the different chemical components and density models. In the n300 model, we find a good correlation between the total density and the H$_2$ density within their error bars. This is because about 98\% of the initial atomic hydrogen gas is in molecular form at this time, meaning that H$_2$ is an excellent tracer of the total density. Furthermore, $^{13}$CO is optically thin and thus we expect the $^{13}$CO intensity PPV model to trace the underlying 3D PPP CO density model well, which we can see in Figure \ref{fig:CVISF}. This argument is also valid for the other density models. On the other hand, $^{12}$CO is optically thick and traces H$_2$ well only in dense regions \citep[see][]{BertramEtAl2014}. This is why it lies between the total density and the CO density models. We discuss the physical implications for this result in Sec. \ref{subsec:fractaldim}.

Turning to the low-density model n30, we find a larger discrepancy in the slopes of the H$_2$ density and the total density field, because only 61\% of the initial atomic hydrogen gas is now in molecular form. Thus, the H$_2$ gas does not trace the total gas very well, although it is still the best tracer for the total density field in general. Looking at the different CO tracers, we find a good correlation between the CO density, $^{12}$CO and $^{13}$CO intensity cases within the error bars.

The n100 model is intermediate between n30 and n300. We find a better correlation of the H$_2$ and the total density cases than in the n30 model, which comes from the higher fractional abundance of H$_2$ compared to the n30 model. Furthermore, we again obtain a good correlation between all CO tracer components within their error bars, which means that the turbulence statistics in all three cases are similar.

Nevertheless, regarding the question of whether CO tracer molecules can be used to infer the dynamics of the underlying total density or H$_2$ density field, we find that in all density models, the slopes of the total density or the H$_2$ density field might be underestimated by a factor up to 3 by the use of different CO tracers.

\subsection{Slopes of the Fourier spectra}
\label{subsec:slopesFS}

\begin{table}
\begin{tabular}{l|c|c|c}
\hline\hline
Model name & n30 & n100 & n300 \\
\hline
Total density & $-3.49 \pm 0.11$ & $-3.57 \pm 0.09$ & $-3.62 \pm 0.09$ \\
H$_2$ density & $-3.41 \pm 0.11$ & $-3.54 \pm 0.09$ & $-3.60 \pm 0.09$ \\
CO density & $-3.04 \pm 0.09$ & $-2.98 \pm 0.07$ & $-3.27 \pm 0.10$ \\
$^{12}$CO intensity & $-3.01 \pm 0.10$ & $-3.01 \pm 0.08$ & $-3.36 \pm 0.08$ \\
$^{13}$CO intensity & $-3.00 \pm 0.10$ & $-2.87 \pm 0.09$ & $-3.30 \pm 0.09$ \\
\hline
\end{tabular}
\caption{Spectral slopes for all chemical components and density models shown in Figure \ref{fig:Spectra}. In order to compare our spectral slope values $\alpha$ to those given in \citet{LazarianEtAl2004}, we have to subtract a value of -2, i.e. $\beta = \alpha - 2$. Independent of the individual component, we observe a saturation of the spectral slopes to a value of $\beta \approx -3$.}
\label{tab:Fourier}
\end{table}

We can understand the significant differences between the H$_2$ and the CO tracer cases in the Fourier spectra of the distinct density models by comparing Figure \ref{fig:CVISF} with Figure \ref{fig:Spectra}. In Eq. \ref{eq:scaling} we have shown that the slopes $\alpha$ of the Fourier spectra are related to the structure functions of the turbulence over the second order structure function $\zeta_2$ via $\alpha = -1 - \zeta_2$. The interpretation of slope differences $\Delta \alpha = \zeta_2^{(b)} - \zeta_2^{(a)}$ in Figure \ref{fig:Spectra} between different chemical tracers is thus related to the differences in slopes of the second order structure functions. If the differences of values for $\zeta_2$ within one density model are large, we also expect large differences to occur in the slopes of the Fourier spectra, which is indeed the case. The ``average'' difference for $\zeta_2$ between H$_2$ and the CO tracers is largest in the n100 model. Hence, we also find the largest slope differences between those two cases in our Fourier spectra for the n100 model. This is why Fourier slopes in Figure \ref{fig:Spectra} of all CO tracers are significantly flatter then those of H$_2$. Since we always find the highest slope values for the total density and H$_2$ density in Figure \ref{fig:CVISF}, we also measure steeper exponents of those components in the Fourier spectra.

However, we have to keep in mind that we have used 3D statistical measures to understand the behaviour of the 2D centroid velocity maps. Nevertheless, since we only want to compare relative scaling behaviours in our models, we can safely use 3D statistical tools of turbulence analysis in order to work out trends in 2D data available for observers.

Furthermore, \citet{LazarianEtAl2004} predicted that the spectral slope should saturate to a value of -3 for optically thick data in the context of integrated intensities, which was confirmed by several other studies in the past (see, e.g. \citealt{StutzkiEtAl1998,StanimirovicAndLazarian2001,LazarianAndPogosyan2006,PadoanEtAl2006,BurkhartEtAl2010}, \citeauthor{BurkhartEtAl2013a}~2013a). We note that in the context of \citet{LazarianEtAl2004}, the power spectra are defined via $E(k) \propto k^2 P(k) \propto k^\beta$. Thus, if we want to compare our results to those found in \citet{LazarianEtAl2004}, we have to subtract a value of -2 in order to compensate for the $k^2$ scaling, i.e. $\beta = \alpha - 2$, where $\alpha$ denotes the Fourier slopes given in Figure \ref{fig:Spectra} and $\beta$ the slopes listed in \citet{LazarianEtAl2004} or \citeauthor{BurkhartEtAl2013a}~2013a. Table \ref{tab:Fourier} gives an overview about the spectral slopes $\beta$ computed from our slope values $\alpha$ given in Figure \ref{fig:Spectra}. Independent of the individual component and the initial density, we also observe a saturation of the spectral slope indices to a value of $\beta \approx -3$ for CO \citep[see also][]{LazarianEtAl2004}. However, we have to keep in mind that our statistics deal with centroid velocities, while previous studies elaborated the dependence of the spectral slopes on integrated intensities. Thus, we expect some modifications as in our case there is a dependency on both density and velocity, which could explain the slight steepening of slope values to higher densities seen in Table \ref{tab:Fourier}.

\subsection{Projection effects}
\label{subsec:projection}

A direct comparison of CVISF to models of 3D turbulence such as proposed by \citet{SheAndLeveque1994} or \citet{BoldyrevEtAl2002} is complicated, since maps of centroid velocities are a complex convolution of the density (or brightness temperature) with the velocity field. Moreover, previous studies have shown that physical effects like opacity or the sonic Mach number also significantly influence the result of the CV analysis (see, e.g. \citealt{LazarianEtAl2004}, \citeauthor{BurkhartEtAl2013a}~2013a and \citealt{BurkhartEtAl2014}). Nevertheless, CVI statistics provide a useful tool to study effects of turbulence in observational measurements and to compare those to numerical experiments, since observations rely on spectral PPV information from which the CV maps can directly be inferred.

As shown by \citet{LazarianAndEsquivel2003}, \citet{OssenkopfEtAl2006}, or \citet{EsquivelEtAl2007}, CVI statistics are significantly influenced by the underlying density field and thus differ from the pure 3D velocity statistics, unless the ratio of the density dispersion to the mean density is small. This is usually the case in subsonic flows where the Mach number is supposed to be small, which is clearly not the case in our simulations, where we measure supersonic Mach numbers. Table \ref{tab:setup} gives both the mean Mach number and the ratio of density dispersion and mean density, $\sigma_{\rho} / \langle \rho \rangle$. We find values $\sigma_{\rho} / \langle \rho \rangle \gtrsim 3$ and therefore expect the mutual convolution of density and velocity field to play a significant role. \citet{OssenkopfEtAl2006} give an upper limit of $\sigma_{\rho} / \langle \rho \rangle \lesssim 0.5$ in order to get a proper matching of the CVISF with the 3D structure functions of the turbulence. We thus expect a significant loss in the signatures of intermittency in our CVI statistics compared to the proper 3D data, which limits a direct comparison of the CVISF to the different moments of the structure functions of the underlying 3D velocity field. For further information about the density variance-Mach number relation we refer the reader to \citet{SemadeniAndGarcia2001}, \citet{LemasterAndStone2008}, \citet{FederrathEtAl2008}, \citet{FederrathEtAl2010}, \citet{MolinaEtAl2012}, \citet{BurkhartEtAl2012} and \citet{KonstandinEtAl2012b}.

\subsection{Connection to the fractal dimension}
\label{subsec:fractaldim}

As shown in Figure \ref{fig:CVISF}, we find increasing differences between the H$_2$ and the CO tracers with higher order $p$ of the CVISF for all density models. Although we analyse 2D CVI statistics, we can nevertheless gain useful information about the most dissipative structures in our simulations using theoretical models of 3D velocity statistics. However, as discussed in \citet{KowalLazarianBeresnyak2007}, we have to keep in mind that supersonic turbulence contaminates the scaling relations of the CV statistics due to a complex convolution of the supersonic density with the velocity.

In the context of \citet{SheAndLeveque1994} ($C = 2$), Eq. \ref{eq:B02} predicts larger slope values $\zeta_p$ than in the model given by \citet{BoldyrevEtAl2002} ($C = 1$), in which $\zeta_p$ becomes flatter for higher order $p$. Thus, the increasing differences between the H$_2$ and the CO tracers with higher order $p$ of the CVISF seem to be related to the fractal co-dimension $C$, which can be interpreted as a space-filling factor, describing the distribution of gas in the MC. As argued in \citet{BertramEtAl2014}, H$_2$ is more extended than CO as it is better able to self-shield in diffuse regions, while CO is a good tracer of H$_2$ only in compact regions. Hence, we would predict a larger fractal co-dimension for the H$_2$ gas than for the various CO tracers, leading to the increasing differences in $\zeta_p$ as observed in Figure \ref{fig:CVISF} between H$_2$ and CO. However, due to projection and resolution effects it is rather complicated to give general values for the fractal co-dimension of the 3D velocity statistics. The question of how the fractal dimensions of 2D and 3D velocity statistics are related with each other can be addressed in a follow-up study.

\subsection{Limitations of the models}
\label{subsec:limitations}

As described in \citet{BertramEtAl2014}, there are some assumptions and limitations inherent to our numerical models that we must keep in mind when interpreting the results. For example, our models do not account for self-gravity, star formation, stellar feedback (e.g. by SN, stellar radiation, etc.) or large-scale dynamics (e.g. spiral arms, SN shells, etc.). Furthermore, we find a minor resolution dependence of the CVISF slopes for CO as discussed in Appendix \ref{app:resolution}, which limits the ability to directly compare our theoretical values with observational measurements. However, we can gain useful information about the underlying physics through the relative scaling of the CVISF slopes to get an idea of how the chemistry and a radiative transfer affect the trends seen in tools like CVI statistics or Fourier spectra, which are commonly used by observers.

For future investigations, higher resolution models would be needed in order to measure converged slope values that can be compared to observations. Additionally, simulations that span a wider range of physical parameters, i.e. different metallicities or radiation fields, are important to study in order to analyse the influence on the CVI statistics. Star formation, feedback and stellar winds could also help to improve our understanding of turbulence in MCs.

\section{Summary and Conclusions}
\label{sec:summary}

We analysed centroid velocity increment structure functions (CVISF) and Fourier spectra of MCs with time-dependent chemistry and a radiative transfer post-processing for models of different initial number densities and chemical components: the total number density, H$_2$ and CO density (each without radiative transfer) as well as $^{12}$CO and $^{12}$CO intensity (both with radiative transfer). In each case, we computed CVISF and their Fourier spectra and analysed the slopes of the CVISF within a fitting range used by \citet{FederrathEtAl2010} and \citet{KonstandinEtAl2012a} in order to study the influence of the chemistry and the radiative transfer on the results of the CVI statistics. We report the following findings:
\begin{itemize}
\item Optical depth effects can be important for the CVI structure analysis. We find different behaviours of the CVISF slopes for $^{12}$CO and $^{13}$CO in different density environments (see Sec. \ref{subsec:slopesCVISF}).
\item We find the slopes of H$_2$ in Fourier spectra generally to be steeper than the slopes of different CO tracers (see Sec. \ref{subsec:slopesFS}). We also find the slopes of the CVISF for the total density and H$_2$ density to be steeper by a factor of up to 3 than the slopes of different CO tracers.
\item We find $\beta \lesssim -3$ for all spectral slope values measured in this study (see Sec. \ref{subsec:slopesFS}), saturating at a value of -3 in the synthetic CO emission data cubes \citep[see also][]{LazarianEtAl2004}.
\item We expect the CO gas to have a significantly smaller fractal co-dimension than the H$_2$ gas, which means that it is less space-filling than H$_2$ (see Sec. \ref{subsec:fractaldim}).
\item We do not find any variations in the slopes of the CVISF using more realistic abundances of $^{13}$CO in MCs compared to a constant $^{12}$CO/$^{13}$CO scaling ratio throughout the whole cloud (see Sec. \ref{subsec:ab13CO}).
\item Following the results of previous studies (see Sec. \ref{subsec:projection}), we expect a significant loss of information in the turbulence statistics due to projection, since we find values of $\sigma_{\rho} / \langle \rho \rangle \gtrsim 3$ in our simulations of supersonic turbulence. This limits a direct test of the CVISF to the different moments of the 3D structure functions.
\end{itemize}

\section*{Acknowledgements}

We thank L\'{a}szl\'{o} Sz{\H u}cs and Frank Bigiel for informative discussions about the project and the referee for a timely and detailed report. EB, LK, RS, SCOG and RSK acknowledge support from the Deutsche Forschungsgemeinschaft (DFG) via the SFB 881 (sub-projects B1, B2, B5 and B8) ``The Milky Way System'', and the SPP (priority program) 1573, ``Physics of the ISM''. Most of the simulations analysed in this paper were performed using the Ranger cluster at the Texas Advanced Computing Center, using time allocated as part of Teragrid project TG-MCA99S024. Additional simulations were performed on the \textit{kolob} cluster at the University of Heidelberg, which is funded in part by the DFG via Emmy-Noether grant BA 3706, and via a Frontier grant of Heidelberg University, sponsored by the German Excellence Initiative as well as the Baden-W\"urttemberg Foundation. RSK acknowledges support from the European Research Council under the European Communitys Seventh Framework Programme (FP7/2007-2013) via the ERC Advanced Grant "STARLIGHT: Formation of the First Stars" (project number 339177).

\bibliographystyle{mn2e}
\bibliography{lit/literature}

\begin{thebibliography}{}

\bibitem[\protect\citeauthoryear{Benzi, Ciliberto, Tripiccione, Baudet,
  Massaioli \& Succi}{Benzi et~al.}{1993}]{BenziEtAl1993}
Benzi R.,  Ciliberto S.,  Tripiccione R.,  Baudet C.,  Massaioli F.,    Succi
  S.,  1993, Phys. Rev. E, 48, R29

\bibitem[\protect\citeauthoryear{{Beresnyak} \& {Lazarian}}{{Beresnyak} \&
  {Lazarian}}{2009}]{BeresnyakAndLazarian2009}
{Beresnyak} A.,  {Lazarian} A.,  2009, \apj, 702, 1190

\bibitem[\protect\citeauthoryear{{Bertram}, {Shetty}, {Glover}, {Klessen},
  {Roman-Duval} \& {Federrath}}{{Bertram} et~al.}{2014}]{BertramEtAl2014}
{Bertram} E.,  {Shetty} R.,  {Glover} S.~C.~O.,  {Klessen} R.~S.,
  {Roman-Duval} J.,    {Federrath} C.,  2014, \mnras, 440, 465

\bibitem[\protect\citeauthoryear{{Boldyrev}, {Nordlund} \& {Padoan}}{{Boldyrev}
  et~al.}{2002}]{BoldyrevEtAl2002}
{Boldyrev} S.,  {Nordlund} {\AA}.,    {Padoan} P.,  2002, Physical Review
  Letters, 89, 031102

\bibitem[\protect\citeauthoryear{{Brunt} \& {Heyer}}{{Brunt} \&
  {Heyer}}{2002}]{BruntAndHeyer2002a}
{Brunt} C.~M.,  {Heyer} M.~H.,  2002, \apj, 566, 276

\bibitem[\protect\citeauthoryear{{Burgers}}{{Burgers}}{1948}]{Burgers1948}
{Burgers} J.~M.,  1948, Adv. Appl. Mech., 1, 171

\bibitem[\protect\citeauthoryear{{Burkhart} \& {Lazarian}}{{Burkhart} \&
  {Lazarian}}{2012}]{BurkhartEtAl2012}
{Burkhart} B.,  {Lazarian} A.,  2012, \apjl, 755, L19

\bibitem[\protect\citeauthoryear{{Burkhart}, {Lazarian}, {Le{\~a}o}, {de
  Medeiros} \& {Esquivel}}{{Burkhart} et~al.}{2014}]{BurkhartEtAl2014}
{Burkhart} B.,  {Lazarian} A.,  {Le{\~a}o} I.~C.,  {de Medeiros} J.~R.,
  {Esquivel} A.,  2014, \apj, 790, 130

\bibitem[\protect\citeauthoryear{{Burkhart}, {Lazarian}, {Ossenkopf} \&
  {Stutzki}}{{Burkhart} et~al.}{013a}]{BurkhartEtAl2013a}
{Burkhart} B.,  {Lazarian} A.,  {Ossenkopf} V.,    {Stutzki} J.,  {2013a},
  \apj, 771, 123

\bibitem[\protect\citeauthoryear{{Burkhart}, {Ossenkopf}, {Lazarian} \&
  {Stutzki}}{{Burkhart} et~al.}{013b}]{BurkhartEtAl2013b}
{Burkhart} B.,  {Ossenkopf} V.,  {Lazarian} A.,    {Stutzki} J.,  {2013b},
  \apj, 771, 122

\bibitem[\protect\citeauthoryear{{Burkhart}, {Stanimirovi{\'c}}, {Lazarian} \&
  {Kowal}}{{Burkhart} et~al.}{2010}]{BurkhartEtAl2010}
{Burkhart} B.,  {Stanimirovi{\'c}} S.,  {Lazarian} A.,    {Kowal} G.,  2010,
  \apj, 708, 1204

\bibitem[\protect\citeauthoryear{Cho, Lazarian \& Vishniac}{Cho
  et~al.}{2002}]{ChoLazarianVishniac2002}
Cho J.,  Lazarian A.,    Vishniac E.~T.,  2002, The Astrophysical Journal
  Letters, 566, L49

\bibitem[\protect\citeauthoryear{{Dobler}, {Haugen}, {Yousef} \&
  {Brandenburg}}{{Dobler} et~al.}{2003}]{DoblerEtAl2003}
{Dobler} W.,  {Haugen} N.~E.,  {Yousef} T.~A.,    {Brandenburg} A.,  2003,
  \physreve, 68, 026304

\bibitem[\protect\citeauthoryear{{Draine}}{{Draine}}{1978}]{Draine1978}
{Draine} B.~T.,  1978, \apjs, 36, 595

\bibitem[\protect\citeauthoryear{Dubrulle}{Dubrulle}{1994}]{Dubrulle1994}
Dubrulle B.,  1994, Phys. Rev. Lett., 73, 959

\bibitem[\protect\citeauthoryear{{Dullemond}}{{Dullemond}}{2012}]{Dullemond201%
2}
{Dullemond} C.~P.,  2012, \apscl

\bibitem[\protect\citeauthoryear{{Elmegreen} \& {Scalo}}{{Elmegreen} \&
  {Scalo}}{2004}]{ElmegreenAndScalo2004}
{Elmegreen} B.~G.,  {Scalo} J.,  2004, \araa, 42, 211

\bibitem[\protect\citeauthoryear{{Esquivel}, {Lazarian}, {Horibe}, {Cho},
  {Ossenkopf} \& {Stutzki}}{{Esquivel} et~al.}{2007}]{EsquivelEtAl2007}
{Esquivel} A.,  {Lazarian} A.,  {Horibe} S.,  {Cho} J.,  {Ossenkopf} V.,
  {Stutzki} J.,  2007, \mnras, 381, 1733

\bibitem[\protect\citeauthoryear{{Falkovich}}{{Falkovich}}{1994}]{Falkovich199%
4}
{Falkovich} G.,  1994, Physics of Fluids, 6, 1411

\bibitem[\protect\citeauthoryear{{Federrath}, {Klessen} \&
  {Schmidt}}{{Federrath} et~al.}{2008}]{FederrathEtAl2008}
{Federrath} C.,  {Klessen} R.~S.,    {Schmidt} W.,  2008, \apjl, 688, L79

\bibitem[\protect\citeauthoryear{{Federrath}, {Roman-Duval}, {Klessen},
  {Schmidt} \& {Mac Low}}{{Federrath} et~al.}{2010}]{FederrathEtAl2010}
{Federrath} C.,  {Roman-Duval} J.,  {Klessen} R.~S.,  {Schmidt} W.,    {Mac
  Low} M.-M.,  2010, \aap, 512, A81

\bibitem[\protect\citeauthoryear{Frisch \& Kolmogorov}{Frisch \&
  Kolmogorov}{1995}]{Frisch1995}
Frisch U.,  Kolmogorov A.,  1995, Turbulence: The Legacy of A.N. Kolmogorov.
Cambridge University Press

\bibitem[\protect\citeauthoryear{{Glover} \& {Clark}}{{Glover} \&
  {Clark}}{2012}]{GloverAndClark2012}
{Glover} S.~C.~O.,  {Clark} P.~C.,  2012, \mnras, 421, 116

\bibitem[\protect\citeauthoryear{{Glover}, {Federrath}, {Mac Low} \&
  {Klessen}}{{Glover} et~al.}{2010}]{GloverEtAl2010}
{Glover} S.~C.~O.,  {Federrath} C.,  {Mac Low} M.-M.,    {Klessen} R.~S.,
  2010, \mnras, 404, 2

\bibitem[\protect\citeauthoryear{{Glover} \& {Mac Low}}{{Glover} \& {Mac
  Low}}{2007}]{GloverAndMacLow2007}
{Glover} S.~C.~O.,  {Mac Low} M.-M.,  2007, \apjs, 169, 239

\bibitem[\protect\citeauthoryear{{Glover} \& {Mac Low}}{{Glover} \& {Mac
  Low}}{2011}]{GloverAndMacLow2011}
{Glover} S.~C.~O.,  {Mac Low} M.-M.,  2011, \mnras, 412, 337

\bibitem[\protect\citeauthoryear{{Goldreich} \& {Sridhar}}{{Goldreich} \&
  {Sridhar}}{1995}]{GoldreichAndSridhar1995}
{Goldreich} P.,  {Sridhar} S.,  1995, \apj, 438, 763

\bibitem[\protect\citeauthoryear{{Haugen} \& {Brandenburg}}{{Haugen} \&
  {Brandenburg}}{2004}]{HaugenAndBrandenburg2004}
{Haugen} N.~E.~L.,  {Brandenburg} A.,  2004, \physreve, 70, 026405

\bibitem[\protect\citeauthoryear{{Hayes}, {Norman}, {Fiedler}, {Bordner}, {Li},
  {Clark}, {ud-Doula} \& {Mac Low}}{{Hayes} et~al.}{2006}]{HayesEtAl2006}
{Hayes} J.~C.,  {Norman} M.~L.,  {Fiedler} R.~A.,  {Bordner} J.~O.,  {Li}
  P.~S.,  {Clark} S.~E.,  {ud-Doula} A.,    {Mac Low} M.-M.,  2006, \apjs, 165,
  188

\bibitem[\protect\citeauthoryear{{Hennebelle} \& {Falgarone}}{{Hennebelle} \&
  {Falgarone}}{2012}]{HennebelleAndFalgarone2012}
{Hennebelle} P.,  {Falgarone} E.,  2012, \aapr, 20, 55

\bibitem[\protect\citeauthoryear{{Heyer} \& {Schloerb}}{{Heyer} \&
  {Schloerb}}{1997}]{HeyerAndSchloerb1997}
{Heyer} M.~H.,  {Schloerb} F.~P.,  1997, \apj, 475, 173

\bibitem[\protect\citeauthoryear{{Hily-Blant}, {Falgarone} \&
  {Pety}}{{Hily-Blant} et~al.}{2008}]{Hily-BlantEtAl2008}
{Hily-Blant} P.,  {Falgarone} E.,    {Pety} J.,  2008, \aap, 481, 367

\bibitem[\protect\citeauthoryear{{Kaneda}, {Ishihara}, {Yokokawa}, {Itakura} \&
  {Uno}}{{Kaneda} et~al.}{2003}]{KanedaEtAl2003}
{Kaneda} Y.,  {Ishihara} T.,  {Yokokawa} M.,  {Itakura} K.,    {Uno} A.,  2003,
  Physics of Fluids, 15, L21

\bibitem[\protect\citeauthoryear{{Kolmogorov}}{{Kolmogorov}}{1941}]{Kolmogorov%
1941}
{Kolmogorov} A.,  1941, Akademiia Nauk SSSR Doklady, 30, 301

\bibitem[\protect\citeauthoryear{{Konstandin}, {Federrath}, {Klessen} \&
  {Schmidt}}{{Konstandin} et~al.}{2012}]{KonstandinEtAl2012a}
{Konstandin} L.,  {Federrath} C.,  {Klessen} R.~S.,    {Schmidt} W.,  2012,
  Journal of Fluid Mechanics, 692, 183

\bibitem[\protect\citeauthoryear{{Konstandin}, {Girichidis}, {Federrath} \&
  {Klessen}}{{Konstandin} et~al.}{2012}]{KonstandinEtAl2012b}
{Konstandin} L.,  {Girichidis} P.,  {Federrath} C.,    {Klessen} R.~S.,  2012,
  \apj, 761, 149

\bibitem[\protect\citeauthoryear{{Kowal}, {Lazarian} \& {Beresnyak}}{{Kowal}
  et~al.}{2007}]{KowalLazarianBeresnyak2007}
{Kowal} G.,  {Lazarian} A.,    {Beresnyak} A.,  2007, \apj, 658, 423

\bibitem[\protect\citeauthoryear{{Kritsuk}, {Norman}, {Padoan} \&
  {Wagner}}{{Kritsuk} et~al.}{2007}]{Kritsuk2007}
{Kritsuk} A.~G.,  {Norman} M.~L.,  {Padoan} P.,    {Wagner} R.,  2007, \apj,
  665, 416

\bibitem[\protect\citeauthoryear{{Lazarian} \& {Esquivel}}{{Lazarian} \&
  {Esquivel}}{2003}]{LazarianAndEsquivel2003}
{Lazarian} A.,  {Esquivel} A.,  2003, \apjl, 592, L37

\bibitem[\protect\citeauthoryear{{Lazarian} \& {Pogosyan}}{{Lazarian} \&
  {Pogosyan}}{2000}]{LazarianEtAl2000}
{Lazarian} A.,  {Pogosyan} D.,  2000, \apj, 537, 720

\bibitem[\protect\citeauthoryear{{Lazarian} \& {Pogosyan}}{{Lazarian} \&
  {Pogosyan}}{2004}]{LazarianEtAl2004}
{Lazarian} A.,  {Pogosyan} D.,  2004, \apj, 616, 943

\bibitem[\protect\citeauthoryear{{Lazarian} \& {Pogosyan}}{{Lazarian} \&
  {Pogosyan}}{2006}]{LazarianAndPogosyan2006}
{Lazarian} A.,  {Pogosyan} D.,  2006, \apj, 652, 1348

\bibitem[\protect\citeauthoryear{{Lemaster} \& {Stone}}{{Lemaster} \&
  {Stone}}{2008}]{LemasterAndStone2008}
{Lemaster} M.~N.,  {Stone} J.~M.,  2008, \apjl, 682, L97

\bibitem[\protect\citeauthoryear{{Lis}, {Pety}, {Phillips} \&
  {Falgarone}}{{Lis} et~al.}{1996}]{LisEtAl1996}
{Lis} D.~C.,  {Pety} J.,  {Phillips} T.~G.,    {Falgarone} E.,  1996, \apj,
  463, 623

\bibitem[\protect\citeauthoryear{{Mac Low}}{{Mac Low}}{1999}]{MacLow1999}
{Mac Low} M.-M.,  1999, \apj, 524, 169

\bibitem[\protect\citeauthoryear{{Mac Low} \& {Klessen}}{{Mac Low} \&
  {Klessen}}{2004}]{MacLowAndKlessen2004}
{Mac Low} M.-M.,  {Klessen} R.~S.,  2004, Reviews of Modern Physics, 76, 125

\bibitem[\protect\citeauthoryear{{Mac Low}, {Klessen}, {Burkert} \&
  {Smith}}{{Mac Low} et~al.}{1998}]{MacLowKlessenBurkertSmith1998}
{Mac Low} M.-M.,  {Klessen} R.~S.,  {Burkert} A.,    {Smith} M.~D.,  1998,
  Physical Review Letters, 80, 2754

\bibitem[\protect\citeauthoryear{{McKee} \& {Ostriker}}{{McKee} \&
  {Ostriker}}{2007}]{McKeeAndOstriker2007}
{McKee} C.~F.,  {Ostriker} E.~C.,  2007, \araa, 45, 565

\bibitem[\protect\citeauthoryear{{Molina}, {Glover}, {Federrath} \&
  {Klessen}}{{Molina} et~al.}{2012}]{MolinaEtAl2012}
{Molina} F.~Z.,  {Glover} S.~C.~O.,  {Federrath} C.,    {Klessen} R.~S.,  2012,
  \mnras, 423, 2680

\bibitem[\protect\citeauthoryear{{Nelson} \& {Langer}}{{Nelson} \&
  {Langer}}{1999}]{NelsonAndLanger1999}
{Nelson} R.~P.,  {Langer} W.~D.,  1999, \apj, 524, 923

\bibitem[\protect\citeauthoryear{{Norman}}{{Norman}}{2000}]{Norman2000}
{Norman} M.~L.,  2000, in {Arthur} S.~J.,  {Brickhouse} N.~S.,   {Franco} J.,
  eds, Revista Mexicana de Astronomia y Astrofisica Conference Series Vol.~9 of
  Revista Mexicana de Astronomia y Astrofisica, vol. 27, {Introducing ZEUS-MP:
  A 3D, Parallel, Multiphysics Code for Astrophysical Fluid Dynamics}.
pp 66--71

\bibitem[\protect\citeauthoryear{{Ossenkopf}, {Esquivel}, {Lazarian} \&
  {Stutzki}}{{Ossenkopf} et~al.}{2006}]{OssenkopfEtAl2006}
{Ossenkopf} V.,  {Esquivel} A.,  {Lazarian} A.,    {Stutzki} J.,  2006, \aap,
  452, 223

\bibitem[\protect\citeauthoryear{{Ossenkopf}, {Krips} \& {Stutzki}}{{Ossenkopf}
  et~al.}{2008a}]{OssenkopfEtAl2008a}
{Ossenkopf} V.,  {Krips} M.,    {Stutzki} J.,  2008a, \aap, 485, 917

\bibitem[\protect\citeauthoryear{{Ossenkopf}, {Krips} \& {Stutzki}}{{Ossenkopf}
  et~al.}{2008b}]{OssenkopfEtAl2008b}
{Ossenkopf} V.,  {Krips} M.,    {Stutzki} J.,  2008b, \aap, 485, 719

\bibitem[\protect\citeauthoryear{{Padoan}, {Juvela}, {Kritsuk} \&
  {Norman}}{{Padoan} et~al.}{2006}]{PadoanEtAl2006}
{Padoan} P.,  {Juvela} M.,  {Kritsuk} A.,    {Norman} M.~L.,  2006, \apjl, 653,
  L125

\bibitem[\protect\citeauthoryear{{R{\"o}llig} \& {Ossenkopf}}{{R{\"o}llig} \&
  {Ossenkopf}}{2013}]{RoelligAndOssenkopf2013}
{R{\"o}llig} M.,  {Ossenkopf} V.,  2013, \aap, 550, A56

\bibitem[\protect\citeauthoryear{{Roman-Duval}, {Jackson}, {Heyer}, {Rathborne}
  \& {Simon}}{{Roman-Duval} et~al.}{2010}]{RomanDuvalEtAl2010}
{Roman-Duval} J.,  {Jackson} J.~M.,  {Heyer} M.,  {Rathborne} J.,    {Simon}
  R.,  2010, \apj, 723, 492

\bibitem[\protect\citeauthoryear{{Rosolowsky}, {Goodman}, {Wilner} \&
  {Williams}}{{Rosolowsky} et~al.}{1999}]{RosolowskyEtAl1999}
{Rosolowsky} E.~W.,  {Goodman} A.~A.,  {Wilner} D.~J.,    {Williams} J.~P.,
  1999, \apj, 524, 887

\bibitem[\protect\citeauthoryear{{Scalo} \& {Elmegreen}}{{Scalo} \&
  {Elmegreen}}{2004}]{ScaloAndElmegreen2004}
{Scalo} J.,  {Elmegreen} B.~G.,  2004, \araa, 42, 275

\bibitem[\protect\citeauthoryear{She \& Leveque}{She \&
  Leveque}{1994}]{SheAndLeveque1994}
She Z.-S.,  Leveque E.,  1994, Phys. Rev. Lett., 72, 336

\bibitem[\protect\citeauthoryear{{Shetty}, {Glover}, {Dullemond} \&
  {Klessen}}{{Shetty} et~al.}{011a}]{ShettyEtAl2011a}
{Shetty} R.,  {Glover} S.~C.,  {Dullemond} C.~P.,    {Klessen} R.~S.,  2011a,
  \mnras, 412, 1686

\bibitem[\protect\citeauthoryear{{Sobolev}}{{Sobolev}}{1957}]{Sobolev1957}
{Sobolev} V.~V.,  1957, \sovast, 1, 678

\bibitem[\protect\citeauthoryear{{Stanimirovi{\'c}} \&
  {Lazarian}}{{Stanimirovi{\'c}} \&
  {Lazarian}}{2001}]{StanimirovicAndLazarian2001}
{Stanimirovi{\'c}} S.,  {Lazarian} A.,  2001, \apjl, 551, L53

\bibitem[\protect\citeauthoryear{{Stutzki}, {Bensch}, {Heithausen}, {Ossenkopf}
  \& {Zielinsky}}{{Stutzki} et~al.}{1998}]{StutzkiEtAl1998}
{Stutzki} J.,  {Bensch} F.,  {Heithausen} A.,  {Ossenkopf} V.,    {Zielinsky}
  M.,  1998, \aap, 336, 697

\bibitem[\protect\citeauthoryear{{Sz{\H u}cs}, {Glover} \& {Klessen}}{{Sz{\H
  u}cs} et~al.}{2014}]{SzucsEtAl2014}
{Sz{\H u}cs} L.,  {Glover} S.~C.~O.,    {Klessen} R.~S.,  2014, ArXiv e-prints,
  arXiv:1403.4912

\bibitem[\protect\citeauthoryear{{V{\'a}zquez-Semadeni} \&
  {Garc{\'{\i}}a}}{{V{\'a}zquez-Semadeni} \&
  {Garc{\'{\i}}a}}{2001}]{SemadeniAndGarcia2001}
{V{\'a}zquez-Semadeni} E.,  {Garc{\'{\i}}a} N.,  2001, \apj, 557, 727

\bibitem[\protect\citeauthoryear{{Visser}, {van Dishoeck} \& {Black}}{{Visser}
  et~al.}{2009}]{VisserEtAl2009}
{Visser} R.,  {van Dishoeck} E.~F.,    {Black} J.~H.,  2009, \aap, 503, 323

\bibitem[\protect\citeauthoryear{{Watson}, {Anicich} \& {Huntress}
  Jr.}{{Watson} et~al.}{1976}]{WatsonEtAl1976}
{Watson} W.~D.,  {Anicich} V.~G.,    {Huntress} Jr. W.~T.,  1976, \apjl, 205,
  L165

\end{thebibliography}

\begin{appendix}

\section{Resolution study}
\label{app:resolution}

\begin{figure}
\centerline{
\includegraphics[height=0.70\linewidth,width=1.0\linewidth]{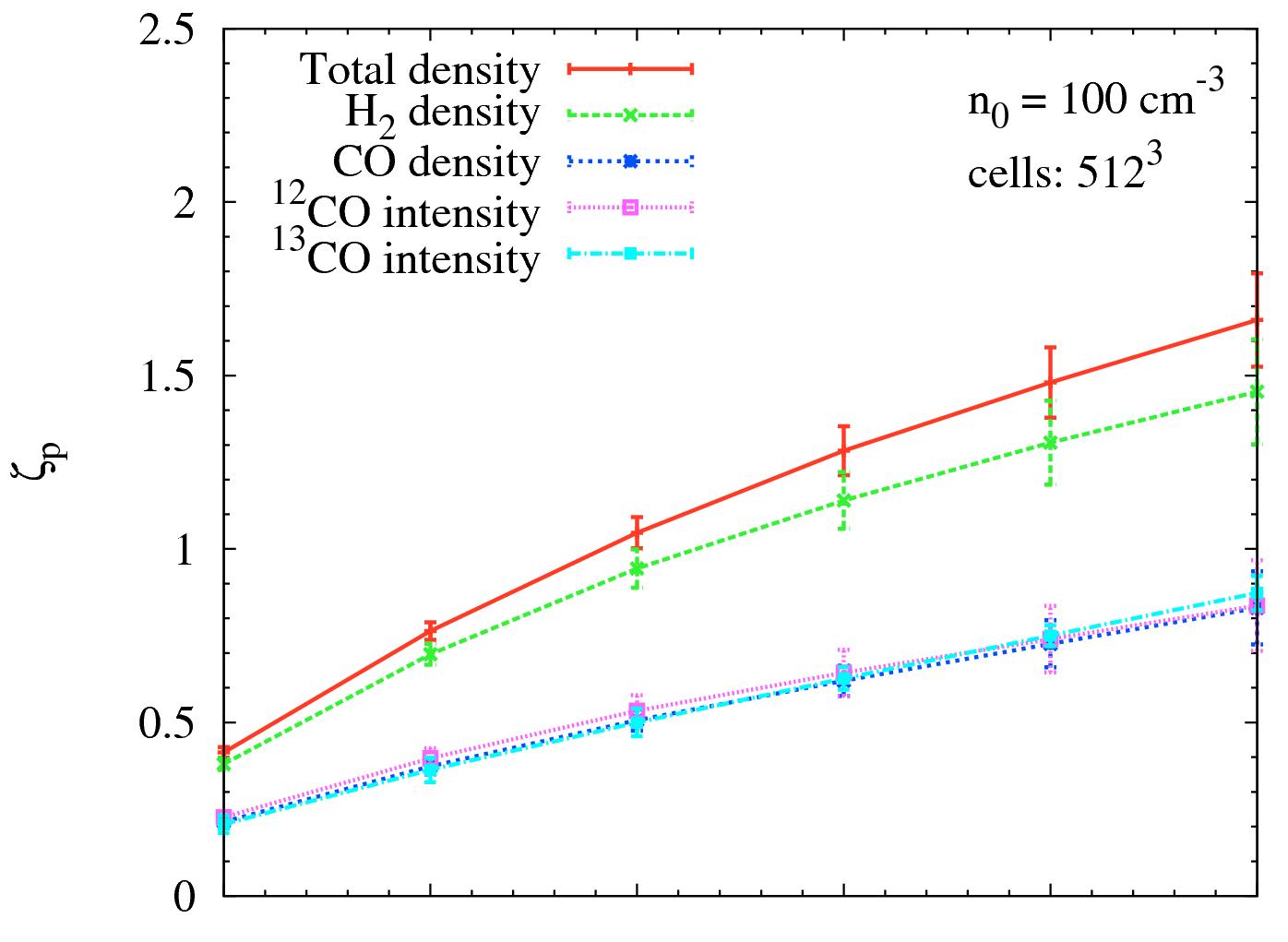} \\
}
\centerline{
\includegraphics[height=0.75\linewidth,width=1.0\linewidth]{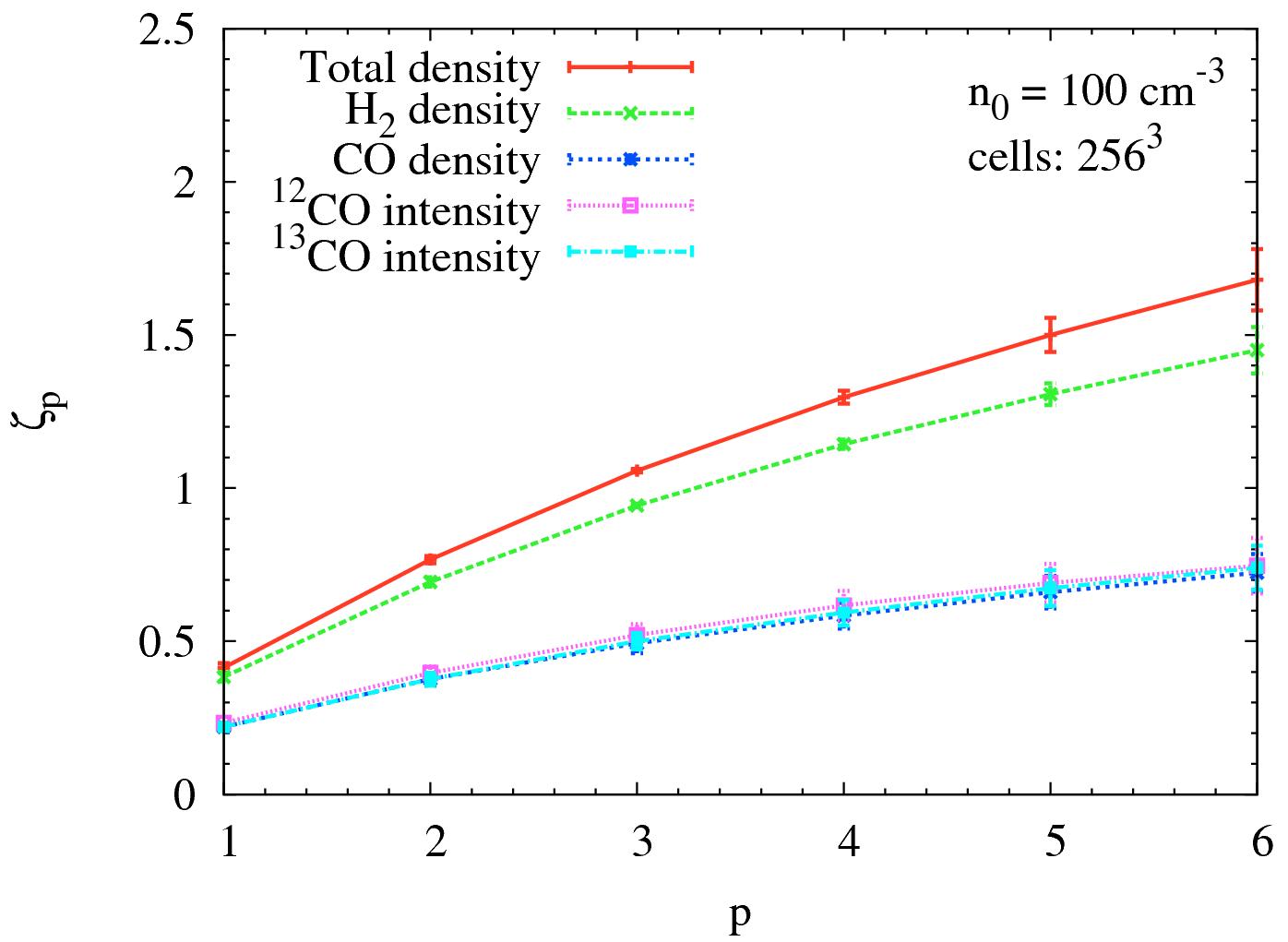} \\
}
\caption{Slopes $\zeta_p$ of the structure functions against order p for all chemical models, i.e. the total density, H$_2$ density and the CO density model as well as the $^{12}$CO and $^{13}$CO intensity model. From top to bottom: runs of different simulations of $512^3$ and $256^3$ grid cells with a fixed initial number density of $n_{0} = 100 \: {\rm cm^{-3}}$. Error bars denote temporal and spatial 1-$\sigma$ fluctuations. Although the individual slope values might be different for the distinct resolutions, the relative scaling behaviour of $\zeta_p$ is conserved.}
\label{fig:res}
\end{figure}

\begin{figure}
\centerline{
\includegraphics[height=0.67\linewidth,width=1.0\linewidth]{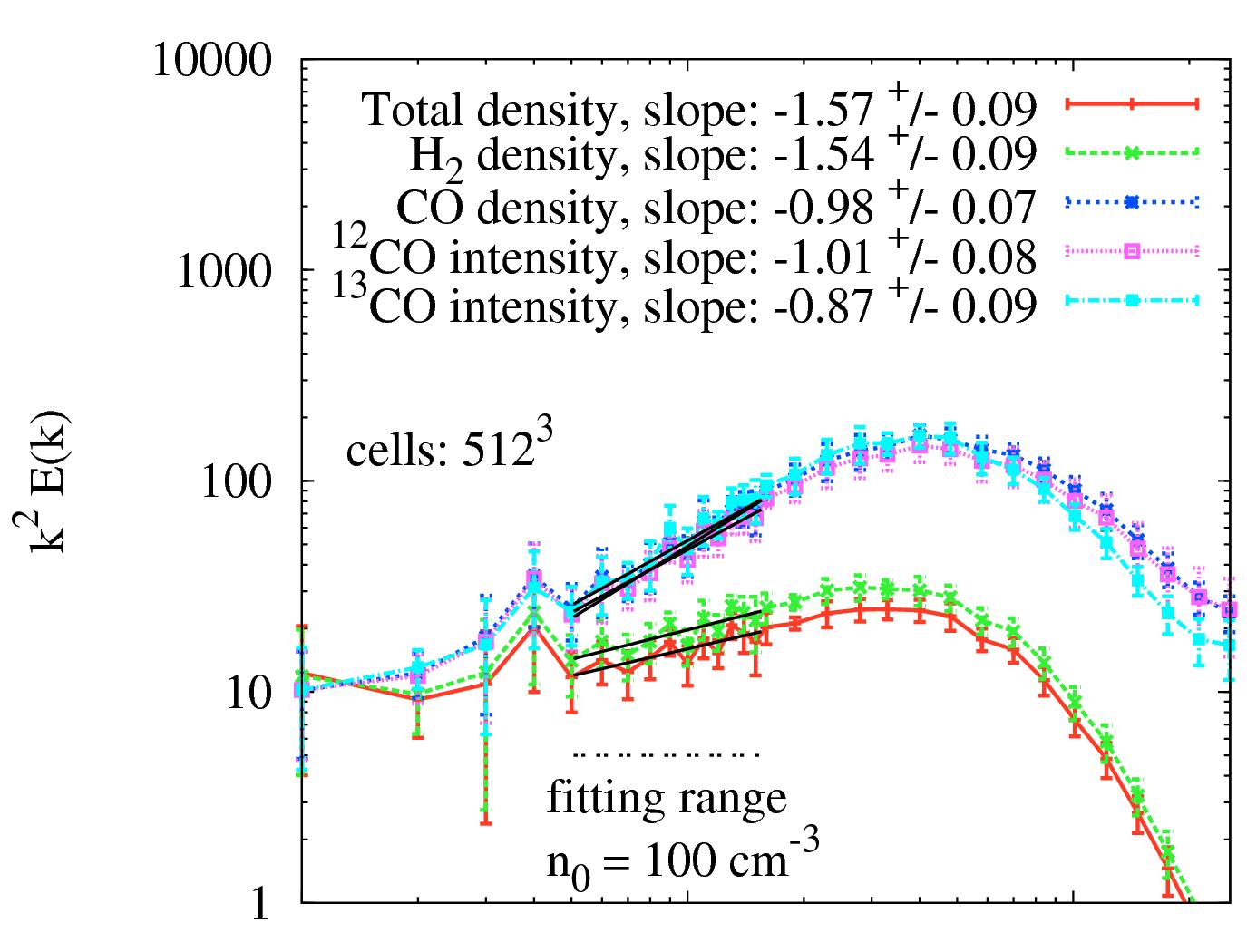} \\
}
\centerline{
\includegraphics[height=0.75\linewidth,width=1.0\linewidth]{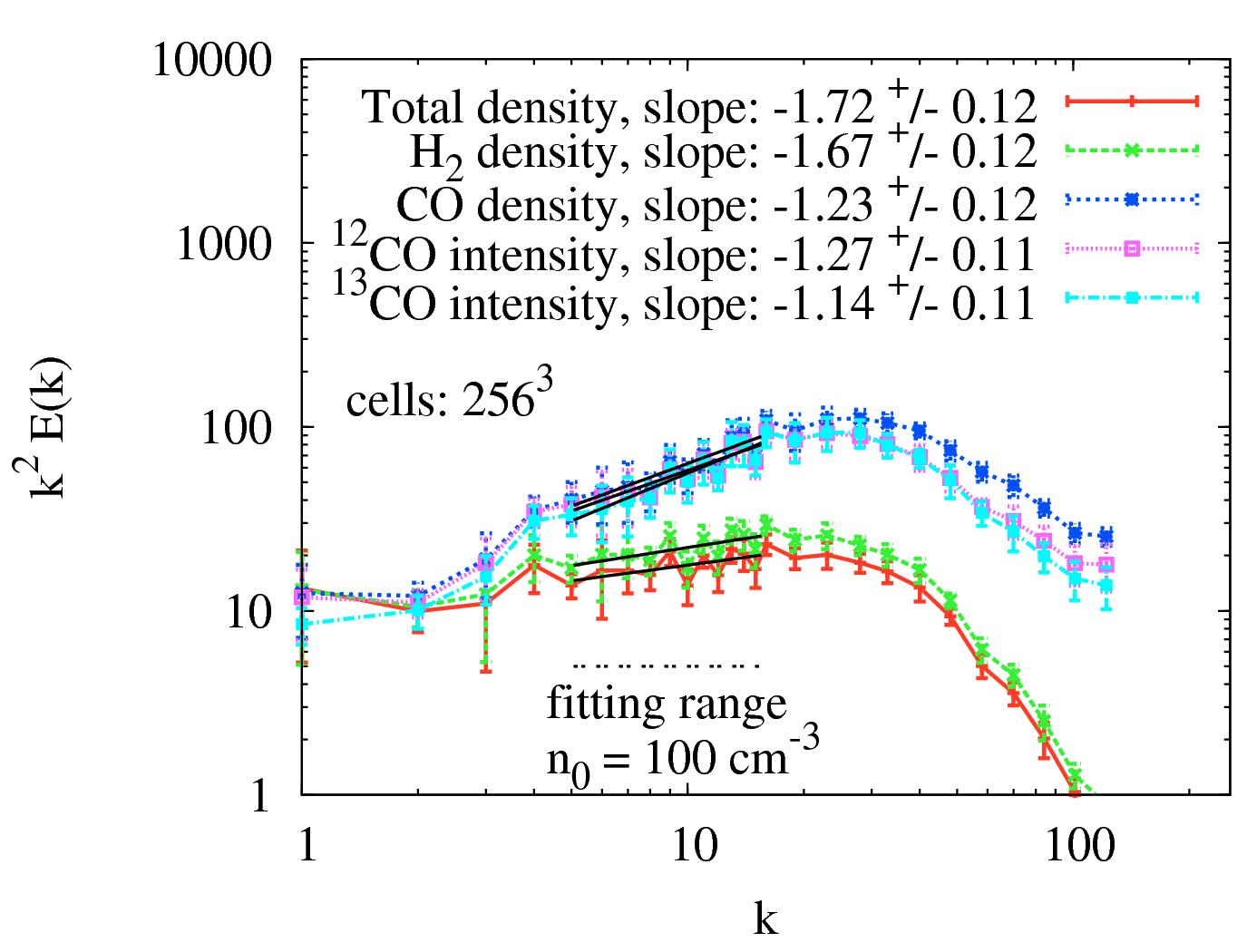} \\
}
\caption{Fourier energy spectra multiplied by $k^2$ as a function of scale $k$ for all chemical components, i.e. the total density, H$_2$ density and the CO density model as well as the $^{12}$CO and $^{13}$CO intensities. From top to bottom: runs of different simulations of $512^3$ and $256^3$ grid cells with a fixed initial number density of $n_{0} = 100 \: {\rm cm^{-3}}$. Error bars denote temporal and spatial 1-$\sigma$ fluctuations. In all models, the total density and H$_2$ density cases show a significantly steeper slope compared to the CO tracer components. The fitting range is indicated by a horizontal dashed line. Slopes with the fitting errors are given in each plot for the different species. Although the individual slope values might be different for the distinct resolutions, the relative scaling behaviour of the energy spectra is conserved.}
\label{fig:resSpectra}
\end{figure}

To study the influence of resolution on the CVISF slopes and the Fourier spectra, we have performed runs of $512^3$ and $256^3$ grid cells and evaluated the CVISF and Fourier slopes for all available chemical components and initial number densities. As an example, Figure \ref{fig:res} shows slopes of the CVISF for an initial number density of $n_0 = 100\,$cm$^{-3}$ for both a resolution of $512^3$ and $256^3$ grid cells. Table \ref{tab:res} gives the corresponding slope values and errors for both resolution models for $p=6$, since we would expect variations of the CVISF slopes due to intermittency particularly at higher orders. For the total density and the H$_2$ density, we measure similar slopes within the errors, while the slopes of the different CO tracers differ by 1\textendash$2\sigma$. This sensitivity to numerical resolution is a consequence of the high degree of chemical inhomogeneity in the numerical simulations. As shown in \citet{BertramEtAl2014}, CO is mainly located in dense gas regions which can be resolved more accurate in the $512^3$ than in the $256^3$ runs. It is thus rather complicated to compare non-converged CVISF slopes with observational measurements. Nevertheless, we find that although the individual slope values might be different for the distinct resolutions, the relative scaling behaviour of $\zeta_p$ is conserved throughout the different resolution runs and hence our fundamental physical conclusions derived in this paper should be unaffected. The same arguments hold for the Fourier spectra given in Figure \ref{fig:resSpectra}. Although the slope values of the individual chemical components are significantly different between the two resolution models, the relative scaling behaviour of the energy spectra is conserved.

\begin{table*}
\begin{tabular}{c||c|c|c|c|c}
\hline\hline
 Resolution & Total density & H$_2$ density & $^{12}$CO density & $^{12}$CO intensity & $^{13}$CO intensity \\
\hline
 $512^3$ & $1.66 \pm 0.13$ & $1.45 \pm 0.15$ & $0.83 \pm 0.10$ & $0.84 \pm 0.13$ & $0.88 \pm 0.05$ \\
 $256^3$ & $1.68 \pm 0.10$ & $1.45 \pm 0.08$ & $0.72 \pm 0.06$ & $0.75 \pm 0.09$ & $0.74 \pm 0.07$ \\
\hline\hline
\end{tabular}
\caption{Slopes of the CVISF for our highest order $p = 6$ for the different chemical components and for different runs with $512^3$ and $256^3$ grid cells. As an example, we show values for a fixed initial number density of $n_0 = 100\,$cm$^{-3}$. The slopes of the total density and H$_2$ density cases are similar within their errors, while the slope values of the different CO tracers significantly deviate with resolution.}
\label{tab:res}
\end{table*}

\end{appendix}
\end{document}